\documentclass[aps,prl,twocolumn,longbibliography,preprintnumbers,amsmath,amssymb,superscriptaddress]{revtex4-1}

\usepackage{graphicx}
\usepackage[T1]{fontenc}
\usepackage[colorlinks,bookmarks=false,citecolor=blue,linkcolor=red,urlcolor=blue]{hyperref}
\usepackage[dvipsnames]{xcolor}
\usepackage{times}

\usepackage{bbm}

\newcommand{\be}{\begin{eqnarray}}
\newcommand{\ee}{\end{eqnarray}}

\newcommand{\nn}{\nonumber } 
\newcommand{\Eqref}[1]{Eq.~\eqref{#1}}

\begin{document}

\author{Daniel D. Scherer}
\affiliation{Niels Bohr Institute, University of Copenhagen, Juliane Maries Vej 30, DK-2100 Copenhagen, Denmark}

\author{Brian M. Andersen}
\affiliation{Niels Bohr Institute, University of Copenhagen, Juliane Maries Vej 30, DK-2100 Copenhagen, Denmark}

\title{Spin-Orbit Coupling and Magnetic Anisotropy in Iron-Based Superconductors}

\begin{abstract}

We determine theoretically the effect of spin-orbit coupling on the magnetic excitation spectrum of itinerant multi-orbital systems, with specific application to iron-based superconductors. Our microscopic model includes a realistic ten-band kinetic Hamiltonian, atomic spin-orbit coupling, and multi-orbital Hubbard interactions. Our results highlight the remarkable variability of the resulting magnetic anisotropy despite constant spin-orbit coupling. At the same time, the magnetic anisotropy exhibits robust universal behavior upon changes in the bandstructure corresponding to different materials of iron-based superconductors. A natural explanation of the observed universality emerges when considering optimal nesting as a resonance phenomenon. Our theory is also of relevance to other itinerant system with spin-orbit coupling and nesting tendencies in the bandstructure.

\end{abstract}

\maketitle

{\it{Introduction.}} The investigation of magnetism in Fe-based superconducting materials (FeSCs) has proven to be a very rich avenue of research~\cite{dai}. Symmetry-distinct magnetic phases have been experimentally identified, both colinear and coplanar~\cite{avci14a,bohmer15a,allred15a,zheng16a,malletta,mallettb,allred16a,meier17}, in agreement with theoretical models~\cite{lorenzana08,eremin,gastiasoro15,scherer16,christensen17}. Recently, it was discovered that distinct colinear phases exhibit completely different orientations of the ordered moments~\cite{wasser15}, pointing to effects from spin-orbit coupling (SOC). The SOC is typically considered weak in the FeSCs, and hence neglected in many theoretical studies. However, recent focus on details of magnetic anisotropies as seen by polarized neutron scattering~\cite{ma,li,song16}, including sizable spin gaps in the ordered states $\sim$15 meV~\cite{dai}, and considerable SOC-induced band splittings of $\sim$10-40 meV~\cite{johnson,watson,borisenko}, have reinvigorated the interest in a detailed understanding of SOC and its role in magnetism and superconductivity of these materials. In addition, obtaining a quantitative description of the magnetic anisotropy has important implications for the general understanding of the magnetism in terms of mainly localized or itinerant electrons~\cite{dai,ma}. Finally, we note that the importance of SOC has recently been highlighted through the experimental report of topological states and Majorana fermions in a certain class of FeSCs~\cite{hongding1,hongding2}.

Experimentally, spin-polarized neutron scattering measurements have mapped out the energy ($\omega$) and temperature ($T$) dependence of the magnetic anisotropy. Below, we denote by $M_a$, $M_b$, and $M_c$ the magnetic scattering polarized along the orthorhombic $a$, $b$, and $c$ axes, respectively. Focusing first on undoped BaFe$_2$As$_2$, in the magnetic state below $T_\mathrm{N}$ the scattering fulfills the hierarchy $M_c>M_b>M_a$. This is in agreement with ${\mathbf{Q}}_{\mathrm{AF}}=(\pi,0,\pi)$ ordered moments aligned antiferromagnetically in the $ab$-plane along the longer $a$ axis, and implies that transverse out-of-plane fluctuations along $c$ are cheaper than in-plane transverse fluctuations in the $b$-direction~\cite{dai,qureshi2012,wang,li}. The results in the paramagnetic (PM) state at $T>T_{\mathrm{N}}$ at ${\mathbf{Q}}_{\mathrm{AF}}$ can be summarized by the following points: 1) The low-energy magnetic response is isotropic $M_c\approx M_b\approx M_a$ at high $T$ but becomes increasingly anisotropic with $M_a> M_c \gtrsim M_b$ as $T$ approaches $T_{\mathrm{N}}$~\cite{qureshi2012,li,luo}. The fact that $M_a$ is largest agrees with the condensation of moments along the $a$ axis below $T_{\mathrm{N}}$. 2) This PM magnetic anisotropy close to $T_{\mathrm{N}}$ is observed only at $\omega \lesssim 6$\,meV~\cite{li}. The doping-dependence of the magnetic anisotropy obtained from electron- and hole-doped BaFe$_2$As$_2$~\cite{qureshi2014,matano,luo,zhang,song16}, NaFeAs~\cite{song13}, and FeSe~\cite{ma} has given rise to the following additional points: 3) Doping of BaFe$_2$As$_2$ tends to enhance the $c$-axis polarized low-energy magnetic fluctuations in the PM phase such that a range exists where $M_c \gtrsim M_a > M_b$. The enhanced susceptibility along $c$ is consistent with the out-of-plane moment orientation of the $C_4$-symmetric magnetic phase observed in Na-doped BaFe$_2$As$_2$~\cite{wasser15}. In the nematic PM phase of FeSe, $M_c$ also dominates the inelastic response~\cite{ma}. 4) At sufficiently large doping (e.g. 15\% Ni in BaFe$_2$As$_2$), the magnetic anisotropy vanishes~\cite{liu}.

The hierarchy of the magnetic susceptibilities, their $\omega$- and $T$-dependence, and their switching as a function of doping has remained an outstanding puzzle, and may naively seem at odds with an atomically defined single-ion spin-orbit-generated magnetic anisotropy. For example, it has been suggested that intervening effects of orbital fluctuations may be at play~\cite{li}. Clearly, it is desirable to acquire a microscopic understanding of the interplay between SOC and electronic interactions in the magnetism of FeSCs. 

Here, within a realistic ten-band description that properly incorporates atomic SOC, we provide a theoretical explanation for the above points 1)-4). We classify the spin-resolved contributions to the particle-hole propagator into different types of excitations. By virtue of SOC, the spin-dependent particle-hole excitations generate a hierarchy in the energy gaps for spin excitations. We propose a general mechanism for the doping-dependence of the resulting magnetic anisotropy that turns out to be determined by the position of the optimal nesting of the band on the energy axis and the dominant orbital content of the participating single-particle states. From that perspective, our study is relevant not just to FeSCs, but any itinerant system with SOC and nested bands. Both the $T$- and $\omega$-dependence of the anisotropy follow essentially from the smallness of the SOC energy-scale together with the enhancement of magnetic scattering close to $T$- or interaction-driven SDW-instabilities.  

{\it{Model.}} Upon inclusion of atomic SOC, the itinerant electron system of the FeSC materials is described by a multiorbital Hubbard Hamiltonian $ H = H_{0} + H_{\mathrm{SOC}}  + H_{\mathrm{int}} $ for the electronic degrees of freedom of the $3d$ shell of iron. The non-interacting part describing the electronic structure consists of a hopping Hamiltonian $H_{0}$ and an atomic SOC $H_{\mathrm{SOC}}$. We define the fermionic operators $ c_{l i \mu \sigma}^{\dagger}$, $c_{l i \mu \sigma}$ to create and destroy, respectively, an electron on sublattice $l $ at site $i$ in orbital $\mu$ with spin polarization $\sigma$. $H_{0}$ is written as
\be
\label{eq:hopping}
H_{0} \!=\!\! \sum_{\sigma}\sum_{l,l^{\prime},i,j}\sum_{\mu,\nu} c_{li \mu \sigma}^{\dagger}\left( t_{li;l^{\prime}j}^{\mu\nu}  - \mu_{0} \delta_{ll^{\prime}}\delta_{ij}\delta_{\mu\nu} \right)c_{l^{\prime}j \nu \sigma}, 
\ee
where hopping matrix elements $t_{li;l^{\prime}j}^{\mu\nu}$ are material specific and the electronic filling is fixed by the chemical potential $\mu_{0}$. The indices $l,l^{\prime} \in \{A,B\}$ denote the 2-Fe sublattices, corresponding to the two inequivalent Fe-sites in the 2-Fe unit cell due to the pnictogen(Pn)/chalcogen(Ch) staggering about the FePn/FeCh plane. The orbital indices $\mu,\nu$ label the five $3d$-orbitals at a given Fe-site. The orbitals of $xz$ and $yz$ symmetry transform to $ - xz $ and $ - yz $ under a glide-plane transformation~\cite{cvetkovic13}. Invariance under the glide-plane transformation thus requires a phase difference of $\pi$ between certain inter-orbital hopping-matrix elements. It is convenient to work in a basis where this phase difference is absorbed in the definition of the local basis for the $xz,yz$-orbitals on $A$ and $B$ sublattices, respectively. For the $A$-sublattice let therefore $\mu,\nu \in \{xz, yz, x^2-y^2, xy, 3z^{2}-r^{2}\}$, while for the $B$ sublattice we take $\mu,\nu \in \{\tilde{xz}, \tilde{yz}, x^2-y^2, xy, 3z^{2}-r^{2}\}$, where $\tilde{xz} = - xz$ and $\tilde{yz} = - yz$. In this `phase-staggered' basis, the atomic SOC Hamiltonian becomes
\be 
H_{\mathrm{SOC}} = \frac{\lambda}{2}\sum_{l,i} \sum_{\mu,\nu} \sum_{\sigma,\sigma^{\prime}}
c_{l i \mu \sigma}^{\dagger} [{\bf L}_{l}]_{\mu\nu}\cdot{\boldsymbol \sigma}_{\sigma\sigma^{\prime}}c_{l i \nu \sigma^{\prime}},
\ee
with coupling strength $\lambda$ and the angular momentum operator in vector notation $[{\bf L}_{l}]_{\mu\nu} $ with components $ [L^{x}_{l}]_{\mu\nu}, \, [L^{y}_{l}]_{\mu\nu}, \, [L^{z}_{l}]_{\mu\nu} $ in the phase-staggered basis of 3$d$-orbitals, and $ {\boldsymbol \sigma} $ is the vector of Pauli matrices. The phase-staggering results in different matrix representations on the $A$- and $B$-sublattice for the angular momentum operator. One obtains $ L^{x}_{A} = - L^{x}_{B} $, $ L^{y}_{A} = - L^{y}_{B} $ and $ L^{z}_{A} = L^{z}_{B} $. While the $z$-component remains unaffected, the couplings of $x$- and $y$-components change sign between $A$- and $B$-sublattice. We note that while in the absence of SOC the ten-orbital model is unitarily equivalent to a five-orbital model formulated in the 1-Fe Brillouin zone, the breakdown of this equivalence in the presence of SOC can be understood as due to the two different matrix representations of the angular momentum operator in the phase-staggered basis.
\begin{figure}
\centering
\begin{minipage}{1\columnwidth}
\centering
\includegraphics[width=1\columnwidth]{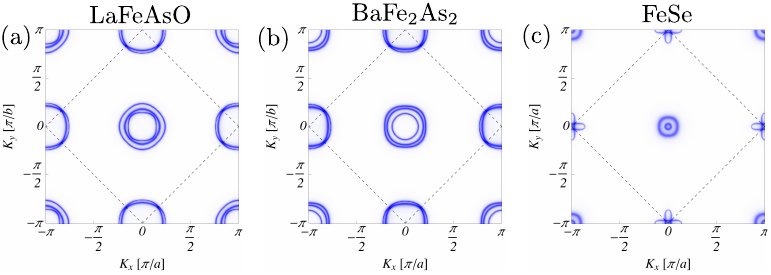}
\end{minipage}
\caption{Fermi surfaces in the 1-Fe BZ ($(K_{x},K_{y})$ denotes momenta in the 1-Fe BZ coordinate system) extracted from the electronic spectral function with $ \mu_{0} = 0 $\,eV and $ \lambda = 0.025 $\,eV for (a) LaFeAsO, (b) BaFe$_{2}$As$_{2}$ and (c) FeSe. The dashed square denotes the 2-Fe BZ.}
\label{fig:FS}
\end{figure}
Electronic interactions of the 3$d$ states are modeled by a local Hubbard-Hund interaction term
\be
\label{eq:interaction}
H_{\mathrm{int}} & = &  
U \sum_{l,i,\mu} n_{l i \mu \uparrow} n_{l i \mu \downarrow} + 
\left(U^{\prime} \!-\! \frac{J}{2}\right) \sum_{l,i,\mu < \nu, \sigma,\sigma^{\prime}} n_{li \mu \sigma} n_{li \nu \sigma^{\prime}} \nn \\
& & \hspace{-2.5em} 
- 2 J \!\! \sum_{l,i, \mu < \nu}{\bf S}_{li\mu}\cdot{\bf S}_{li\nu}  + 
J^{\prime}\!\!\!\! \sum_{l,i, \mu < \nu,\sigma} c_{li\mu\sigma}^{\dagger}c_{li\mu\bar{\sigma}}^{\dagger}c_{li\nu\bar{\sigma}}c_{li\nu\sigma},
\ee
parametrized by an intraorbital Hubbard-$U$, an interorbital coupling $U^{\prime}$, Hund's coupling $J$ and pair hopping $J^{\prime}$, satisfying $U^{\prime} = U - 2J$, $J = J^{\prime}$. The operators for local charge and spin are $n_{li\mu} = n_{li\mu\uparrow} + n_{li\mu\downarrow}$ with $n_{li\mu\sigma} = c_{li\mu\sigma}^{\dagger} c_{li\mu\sigma}$ and ${\bf S}_{li\mu} = 1/2\sum_{\sigma\sigma^{\prime}} c_{li\mu\sigma}^{\dagger} {\boldsymbol \sigma}_{\sigma\sigma^{\prime}}c_{li\mu\sigma^{\prime}}$, respectively.

Below, we will consider three sets of hopping parameters $t_{li;l^{\prime}j}^{\mu\nu}$ for different FeSC parent materials: LaFeAsO~\cite{ikeda10}, BaFe$_{2}$As$_{2}$~\cite{kopernik}, and FeSe~\cite{scherer17}, see Fig.~\ref{fig:FS} for the corresponding Fermi surfaces. The effect of hole- or electron-doping is obtained by a rigid shift in the chemical potential $\mu_{0}$. For further details on the bandstructures and the effects of SOC, we refer to the Supplementary Material (SM)~\cite{sm}.  
%
\begin{figure*}
\centering
\begin{minipage}{1\textwidth}
\centering
\includegraphics[width=1\textwidth]{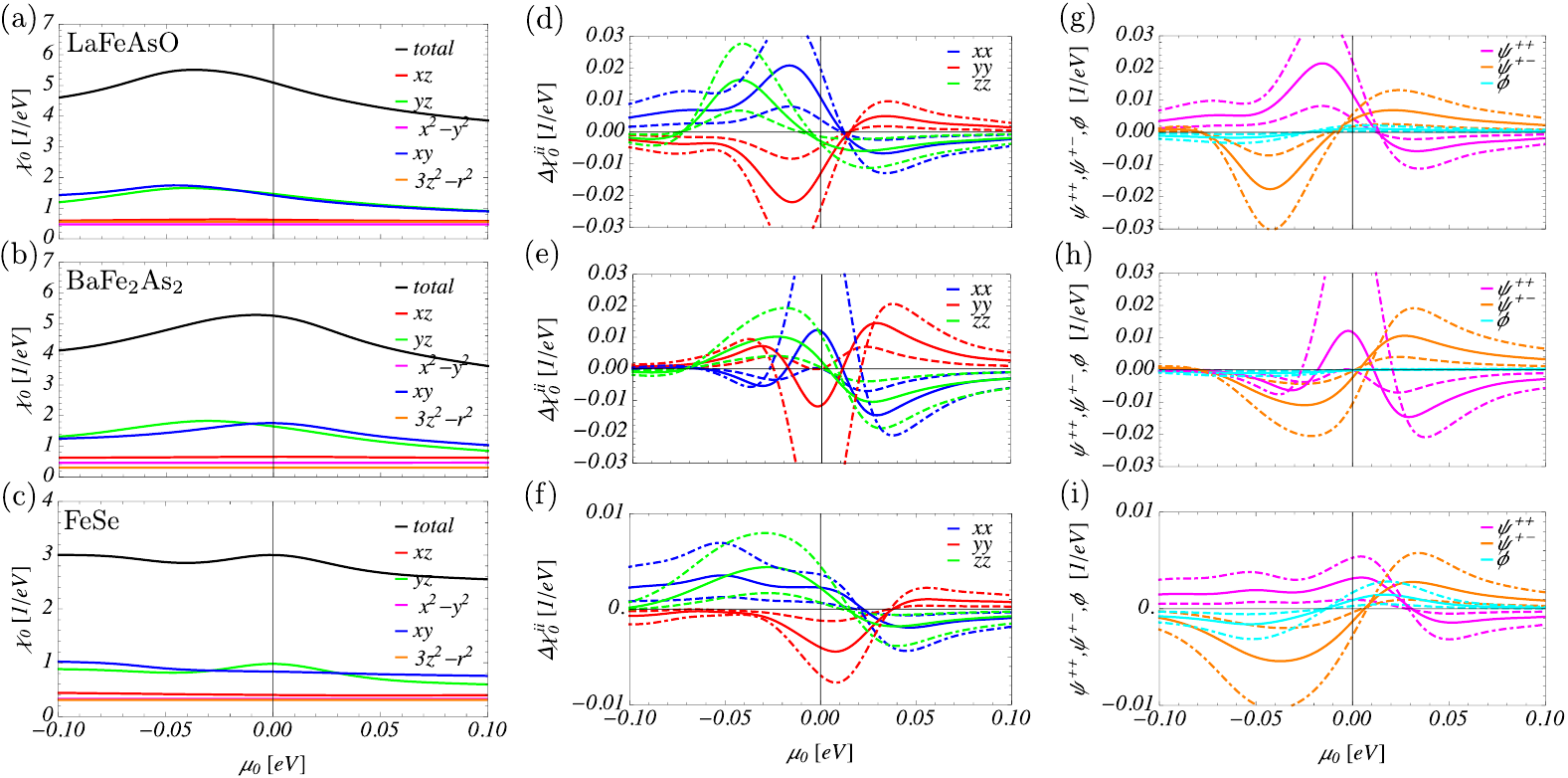}
\end{minipage}
\caption{(a),(b),(c) Chemical potential dependence of the total and orbitally ($\mu = \nu $ only) resolved isotropic contribution to the static non-interacting susceptibility with SOC $ \lambda = 0.025 $\, eV at  $ k_{\mathrm{B}} T = 0.01 $\,eV for the (a) LaFeAsO and (b) BaFe$_{2}$As$_{2}$ and (c) FeSe model with fixed wavevector ${\bf Q}_{1}$. (d),(e),(f) Corresponding anisotropic contributions and (g),(h),(i) summed particle-hole amplitudes contributing to the anisotropic magnetic response for $\lambda = 0.015$\,eV (dashed), $ \lambda = 0.025$\,eV (solid) and $ \lambda = 0.035$\,eV (dot-dashed).}
\label{fig:chi0_chemical_potential}
\end{figure*}
%

{\it{Spin susceptibility.}} To make connection to neutron scattering, we compute the imaginary-time spin-spin correlation function (here $i,j$ refer to the spatial directions $x,y,z$)
\be
\chi^{ij}(\mathrm{i}\omega_n,{\bf q}) = \frac{g^{2}}{2}\int_{0}^{\beta} \! d\tau \,
\mathrm{e}^{\mathrm{i}\omega_n \tau}
\langle \mathcal{T}_{\tau} S^{i}_{{\bf q}}(\tau) S^{j}_{-{\bf q}}(0)\rangle,
\ee
with $g=2$ and the Fourier transformed electron spin operator for the 2-Fe unit cell given as
\be
S^{i}_{{\bf q}}(\tau) = \frac{1}{\sqrt{\mathcal{N}}}\sum_{{\bf k},l,\mu,\sigma,\sigma^{\prime}} c_{{\bf k} - {\bf q}l\mu\sigma}^{\dagger}(\tau) \frac{\sigma_{\sigma\sigma^{\prime}}^{i}}{2}c_{{\bf k}l\mu\sigma^{\prime}}(\tau).
\ee
To account for interaction effects in the weak-coupling regime, we evaluate the correlation functions in the random-phase approximation (RPA) in the absence of spin-rotation invariance, see SM~\cite{sm}. Performing analytic continuation $ \mathrm{i} \omega_{n} \to \omega + \mathrm{i} \eta $, with $ \eta > 0 $ a small smearing parameter, we gain access to the momentum- and frequency-resolved spectral density of magnetic excitations with different spatial polarizations probed by polarized neutron scattering. We have 
\be 
M_{i}(\omega) \sim \mathrm{Im}[\chi^{ii}(\omega + \mathrm{i}\eta,{\bf Q}_{\mathrm{AF}})],
\ee
in a coordinate system $ x = a $, $ y = b $, $ z = c $ aligned with the
orthorhombic crystal axes and $ {\bf Q}_{\mathrm{AF}} = {\bf Q}_{1,2} $ with the nesting vectors $ {\bf Q}_{1} = (\pi,0) $, $ {\bf Q}_{2} = (0,\pi) $, where $ {\bf Q}_{2} $ is related to $ {\bf Q}_{1} $ by a $C_{4}$ rotation in the $ab$ plane. The cross-terms with $ i \neq j $ vanish for the commensurate wavevector $ {\bf Q}_{\mathrm{AF}} $.

Since the interaction term $ H_{\mathrm{int}} $ is rotationally symmetric, it cannot create anisotropy in the magnetic response. Hence, all SOC-driven anisotropy is contained purely in the particle-hole propagator, and therefore the origin of anisotropy is found in the structure of the non-interacting susceptibility. In terms of the sublattice-, orbital-, and spin-resolved electronic Greens function, the non-interacting susceptibility reads
\be 
\label{eq:chi0}
\chi_{0}^{ij}(q) = \frac{1}{4}\sum_{\sigma_{1} \dots \sigma_{4}}\sigma_{\sigma_{1}\sigma_{2}}^{i} \sigma_{\sigma_{3}\sigma_{4}}^{j}G_{\sigma_{2}\sigma_{3}}G_{\sigma_{4}\sigma_{1}},
\ee
where for compact notation we defined 
\be 
G_{\sigma_{2}\sigma_{3}}G_{\sigma_{4}\sigma_{1}} \equiv -\frac{g^{2}}{4 \beta\mathcal{N}} \sum
G_{l\mu\sigma_2;l^{\prime}\nu\sigma_3}(k) 
G_{l^{\prime}\nu\sigma_4;l\mu\sigma_1}(k-q), \nn
\ee
with $ q = (\mathrm{i}\omega_{n},{\bf q}) $ and $ k =  (\mathrm{i}\nu_{p},{\bf k}) $,
$\mathrm{i}\omega_{n},\,\mathrm{i}\nu_{p}$ being bosonic and fermionic Matsubara frequencies, respectively, and the shorthand $ \sum (\dots) = \sum_{k}\sum_{l,l^{\prime}} \sum_{\mu,\nu}  (\dots) $. Performing the Matsubara sum yields a Lindhard-factor dressed by wavevector-dependent matrix elements, see SM~\cite{sm}. We can then extract the isotropic contribution to the susceptibility as
\be 
\chi_{0} = \frac{1}{4} \sum_{\sigma}\left[ 
G_{\sigma\sigma}G_{\sigma\sigma} + 
G_{\sigma\sigma}G_{\bar{\sigma}\bar{\sigma}}
\right].
\ee
The anisotropic contributions, $\Delta \chi_{0}^{ii} = \chi_{0}^{ii} - \chi_{0}$,
can be expressed in terms of three particle-hole amplitudes
\be 
\label{eq:anisotropy1}
\Delta \chi_{0}^{xx} = \psi^{++} -\phi, \quad \Delta \chi_{0}^{yy} = -\psi^{++} - \phi, \\
\label{eq:anisotropy2}
\Delta \chi_{0}^{zz} = -\psi^{+-} + \phi, \hspace{1.5cm}
\ee
where we have defined the summed amplitudes
\be
\label{eq:amplitudes}
\psi^{++} = \frac{1}{2}\sum_{\sigma} G_{\sigma\bar{\sigma}} G_{\sigma\bar{\sigma}}, \quad
\psi^{+-}  = \frac{1}{2}\sum_{\sigma} G_{\sigma\bar{\sigma}}G_{\bar{\sigma}\sigma}, \\
\phi  = \frac{1}{4}\sum_{\sigma}
\left[ 
G_{\sigma\sigma}G_{\sigma\sigma} - 
G_{\sigma\sigma}G_{\bar{\sigma}\bar{\sigma}}
\right]. \hspace{1cm}
\ee
In the non-nematic PM state, the anisotropic response at $ {\bf Q}_{2} $ is related to that at ${\bf Q}_{1}$ by a $C_{4}$ transformation about the $c$-axis: $ \Delta\chi_{0}^{xx/yy}({\bf Q}_{2}) = \Delta\chi_{0}^{yy/xx}({\bf Q}_{1}) $ and $ \Delta\chi_{0}^{zz}({\bf Q}_{2}) = \Delta\chi_{0}^{zz}({\bf Q}_{1}) $. The amplitude $ \phi $, measuring the difference of equal- and opposite-spin (w.r.t. to the $z$-axis pointing out-of-plane) particle-hole propagation is insensitive to a $ C_{4} $ rotation. Likewise, the amplitude $ \psi^{+-} $ corresponds to processes that are possible due to SOC, but do not change the total spin along the $z$-direction. In contrast, the spin-flip amplitude $ \psi^{++} $, where both electron and hole with a fixed initial spin propagate to the opposite spin state by virtue of SOC, reacts by a sign change. A commonality between the bands is the sublattice structure of the anisotropy-generating particle-hole amplitudes. While $\psi^{++}$ receives only inter-sublattice contributions, $\psi^{+-}$ and $\phi$ only come form intra-sublattice terms.

The physical interpretation of the particle-hole bubble diagrams can be made more transparent by considering SOC within perturbation theory. We find that the leading contribution to the anisotropy at $ {\bf Q}_{\mathrm{AF}} $ emerges at order $ \lambda^{2} $ (see SM~\cite{sm} for details). This is in contrast to previous work~\cite{christensen15}, where the leading anisotropy was found to be of the form $J\lambda^{2}$ and depended crucially on a finite Hund's coupling. We additionally investigated the importance of the sign of $\lambda$, see SM~\cite{sm}, and found that results for the magnetic anisotropy are only weakly affected.

{\it{Anisotropy without interactions.}} Our findings for the doping-dependence of the magnetic anisotropy for the non-interacting LaFeAsO, BaFe$_{2}$As$_{2}$, and FeSe models at $k_{\mathrm{B}} T = 0.01$\,eV are shown in Fig.~\ref{fig:chi0_chemical_potential} for several values of $\lambda$. For the 1111 and 122 bands, there exists a clear correlation between the position of the optimal nesting condition on the energy axis (that is only weakly dependent on small $\lambda$), see Fig.~\ref{fig:chi0_chemical_potential}(a),(b), and the central peak in the static anisotropic response as a function of $\mu_{0}$, seen in Fig.~\ref{fig:chi0_chemical_potential}(d),(e). Indeed, the characteristic $\mu_{0}$-dependence of the anisotropy can be qualitatively reproduced in a simple level model, see SM~\cite{sm}, where the optimal nesting condition is replaced by isolated levels with $xy$ and $yz$ orbital content, coupled by SOC. This simple model also provides the same type of spin-dependent particle-hole amplitudes as seen in the tight-binding models, cf. Fig.~\ref{fig:chi0_chemical_potential}(g),(h), pointing to a universal mechanism behind the doping-dependence of the magnetic anisotropy across the FeSC materials. In this picture the behavior of $\Delta\chi_{0}^{ii}$ with doping is determined largely by the position of the optimal nesting condition on the energy axis and the symmetry properties of the participating orbitals. 

For all three tight-binding models, the hierarchy in the magnetic anisotropy changes with $\mu_{0}$. While the different realizations of the hierarchy are already apparent at $ \lambda = 0.015$\,eV, increasing $ \lambda $ enlarges the doping range with a particular form of the hierarchy. For LaFeAsO and BaFe$_{2}$As$_{2}$ we obtain a dominating $ \chi_{0}^{xx} $ in the undoped case, while on the hole-(electron-)doped side, an extended region with dominating $ \chi_{0}^{zz} $ ($\chi_{0}^{yy}$) exists. Sufficiently far away from the nesting resonance, the magnetic anisotropy drops rapidly. These findings are in excellent agreement with properties 3) and 4) highlighted in the introduction. The most prominent difference in the doping-dependence occurs on the hole-doped side, where $\Delta\chi_{0}^{xx}$ and $\Delta\chi_{0}^{yy}$ in the LaFeAsO and FeSe models do not display zero crossings, as opposed to the BaFe$_{2}$As$_{2}$ case. In addition, the FeSe model, where optimal nesting for $xy$ and $yz$ orbitals is weakened and occurs in different places on the $ \mu_{0}$-axis, see Fig.~\ref{fig:chi0_chemical_potential}(c), displays a dominating $ \chi_{0}^{zz} $ in the undoped case for sufficiently large $\lambda$. This agrees with the recent findings in Ref.\onlinecite{ma}, see Fig.~\ref{fig:chi0_chemical_potential}(f). Both weak hole-doping or increasing $ \lambda $ enhance the dominance of out-of-plane spin-fluctuations compared to in-plane fluctuations. The anisotropy is driven by the same type of particle-hole excitations in all models, cf. Fig.~\ref{fig:chi0_chemical_potential}(g),(h),(i). Only $\psi^{++}$ and $\psi^{+-}$ yield sizable contributions in the LaFeAsO and BaFe$_{2}$As$_{2}$ bands, with $\phi$ basically vanishing. For FeSe the $\phi$-amplitude is stronger compared to the 1111 and 122 cases.

{\it{Anisotropy with interactions.}} When including interactions, additional (inter-sublattice and inter-orbital) contributions of the particle-hole propagator enter, that are not included in the susceptibility of the non-interacting system. The properties of the electronic particle-hole propagator together with the interaction vertex, however, fully determine the gap-structure of magnetic excitations with different polarization. While inter-orbital contributions can in principle be enhanced by Hund's coupling, we did not observe a modification of the hierarchy in magnetic anisotropy between the bare and RPA results. In this respect, the static bare susceptibility provides a measure of the gap-sizes of spin excitations with different polarization.

We can thus connect the results in Fig.~\ref{fig:chi0_chemical_potential} to the doping dependence of the magnetic scattering amplitudes $M_{i}$. Focusing on BaFe$_{2}$As$_{2}$, cf. Fig.~\ref{fig:chi0_chemical_potential}(e), our weak-coupling approach yields $ M_{a} > M_{c} > M_{b} $ in an extended region around $\mu_{0}=0$\,eV, consistent with a stripe SDW state with ordered moments along $a$. The formation of a finite SDW order below $ T_{\mathrm{N}} $ results in the gapping of excitations parallel to the moment direction. For sufficiently low $ T $ in the stripe magnetic state, we can thus expect $  M_{c} > M_{b} > M_{a}$. Returning to the discussion of the PM state, for sufficiently strong SOC, hole-doping first leads to a regime with $ M_{c} > M_{a} > M_{b} $, with a subsequent crossover to $ M_{c} > M_{b} > M_{a} $ upon further hole-doping, all consistent with the observed reorientation of magnetic moments in a $C_{4}$-symmetric magnetic phase~\cite{wasser15}.
\begin{figure}[t!]
\centering
\begin{minipage}{1\columnwidth}
\centering
\includegraphics[width=1\columnwidth]{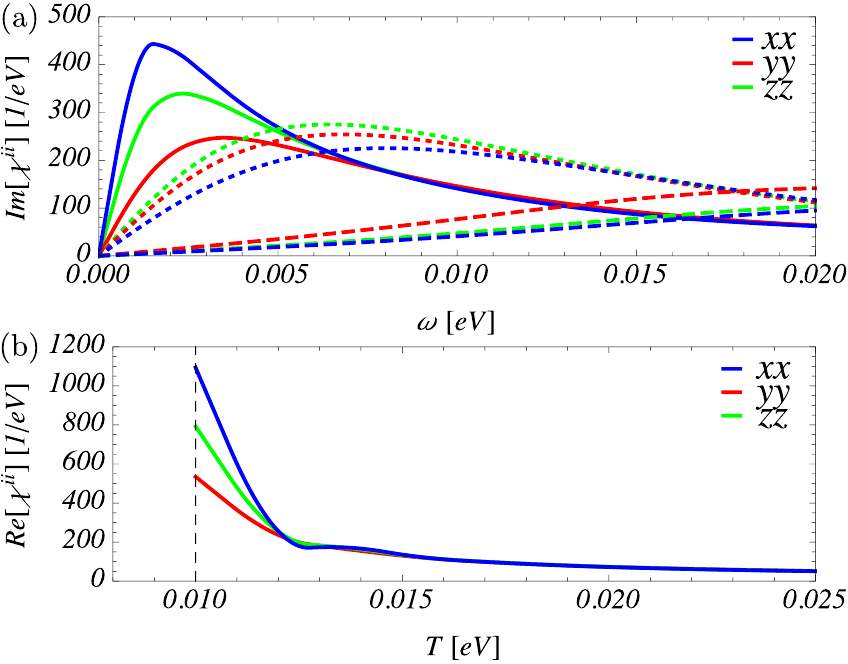}
\end{minipage}
\caption{(a) Imaginary part of the interacting susceptibilities as a function of $\omega$ at wavevector ${\bf Q}_{1}$ for the BaFe$_{2}$As$_{2}$ model with $ \lambda = 0.025 $\,eV at $ k_{\mathrm{B}} T = 0.01 $\, eV close to the interaction driven SDW instability (with $J=U/4$) for different chemical potentials: $\mu_{0} = 0$\,eV, $ U = 0.815 $\,eV (solid), $ \mu_{0} = -0.05$\,eV, $ U = 0.898 $\,eV (dotted) and $ \mu_{0} = 0.05 $\,eV, $ U = 1.030 $\,eV (dashed). (b) $T$-dependence of the static part of the RPA susceptibility for $ \mu = 0 $\,eV with $ \lambda = 0.025 $\,eV and $ U = 0.816 $\,eV, $J=U/4$. The dashed vertical line marks the SDW transition temperature $T_{\mathrm{N}}$.}
\label{fig:chiRPA_omega_temp}
\end{figure}

We show the $\omega$-dependent RPA results for the imaginary part of the susceptibility in the various regimes in Fig.~\ref{fig:chiRPA_omega_temp}(a) for interaction parameters $U$ and $J$ close to the interaction driven SDW-instability with fixed wavevector. For the undoped case ($\mu_{0} = 0$\,eV) the $\omega$-dependent anisotropy in the magnetic scattering is clearly visible and diminishes quickly for $\omega \gtrsim 6-7$\,meV. In the hole- ($\mu = -0.05$\,eV) and electron-doped ($\mu = 0.05$\,eV) cases, the changes in the hierarchy of magnetic scattering can be observed with an overall decrease of the magnetic scattering, while at the same time the anisotropy appears over a larger energy range. These differences to the undoped case are simply due to the increasing degree of incommensurability of the wavevector associated with the leading SDW-instability, while we observe the magnetic scattering at the commensurate wavevector ${\bf Q}_{\mathrm{AF}}$. Thus, the magnetic excitations at ${\bf Q}_{\mathrm{AF}}$ obtain larger gaps for the doped cases than for the undoped case shown in Fig.~\ref{fig:chiRPA_omega_temp}(a). The $T$-dependence of $ \mathrm{Re}[\chi^{ii}(0 + \mathrm{i}\eta,{\bf Q}_{\mathrm{AF}})] $ is shown in Fig.~\ref{fig:chiRPA_omega_temp}(b), where $ \chi^{xx} $ diverges as $ T \to T_{\mathrm{N}} $. The anisotropy increases strongly in the proximity to the SDW transition, while it remains small for elevated $T$. The results shown in Fig.~\ref{fig:chiRPA_omega_temp} are in excellent agreement with the points 1) and 2) discussed in the introduction. Thus, we conclude that the model approach presented here seems to adequately describe the magnetic anisotropy of FeSCs. Interesting future studies include calculations of $\chi^{ii}(q)$ in the presence of SOC in the superconducting state where magnetic anisotropy of the neutron resonance has been reported by polarized neutron scattering~\cite{lipscombe,steffens,luo,CZhang14,wasser,song16,ma}.

\begin{acknowledgments}
We acknowledge discussions with M. H. Christensen, P. Dai,  I. Eremin, A. Kreisel, and F. Lambert, and financial support from the Carlsberg Foundation. 
\end{acknowledgments}




\newpage

\begin{widetext}

\setcounter{equation}{0}
\setcounter{figure}{0}
\setcounter{table}{0}
\setcounter{page}{1}
\makeatletter
\renewcommand{\theequation}{S\arabic{equation}}
\renewcommand{\thefigure}{S\arabic{figure}}
\renewcommand{\bibnumfmt}[1]{[S#1]}
\renewcommand{\citenumfont}[1]{S#1}

\begin{center}
\textbf{\large Supplementary Material: ``Spin-Orbit Coupling and Magnetic Anisotropy in Multiband Metals''}
\end{center}

\section{Effect of $\mathrm{sign}(\lambda)$ on electronic bandstructure and anisotropy}
\label{app:bands}

The spin-orbit coupling (SOC) leads to a non-equivalence of 2-Fe and 1-Fe unit-cell descriptions of the iron-based superconductors (FeSCs). Within the 2-Fe description, a finite SOC splits those states that are degenerate at the boundary of the 2-Fe Brillouin zone due to glide-plane symmetry. The electronic states are, however, still 2-fold degenerate in the paramagnetic state due to time-reversal and inversion symmetry.
\begin{figure*}[h!]
\centering
\begin{minipage}{1\textwidth}
\centering
\includegraphics[width=1\textwidth]{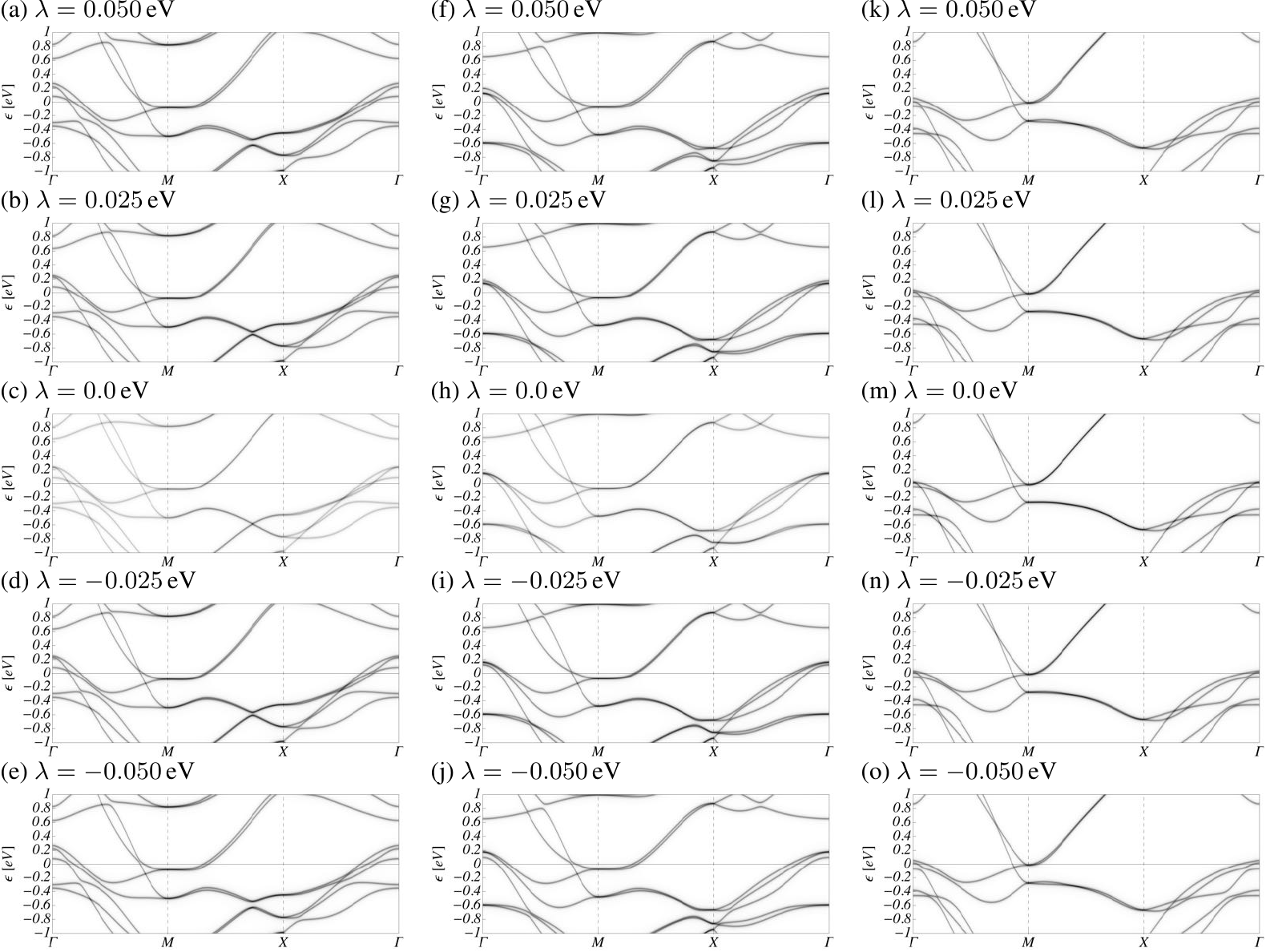}
\end{minipage}
\caption{High-symmetry cuts through bandstructure for (a)-(e) LaFeAsO model~\cite{ikeda2010}, (f)-(j)  BaAs$_{2}$Fe$_{2}$ model~\cite{eschrig2009} and (k)-(o) FeSe model~\cite{scherer2017} with SOC strength $ \lambda $ of varying sign and magnitude. The bandstructures were obtained from the electronic spectral function. The path through momentum space goes from $\Gamma$ to $ M $ over $ X $ and back to $\Gamma$, where momenta are specified with respect to the 2-Fe BZ.}
\label{fig:bands}
\end{figure*}
In Fig.~\ref{fig:bands} we demonstrate the effect of the spin-orbit coupling on the electronic bandstructure for the 2D tight-binding models for LaFeAsO~\cite{ikeda2010}, BaFe$_{2}$As$_{2}$~\cite{eschrig2009} and FeSe~\cite{scherer2017}. The FeSe model was obtained from performing a self-consistent mean-field calculation, yielding a sizable nearest-neighbor hopping renormalization. Within the notation of Ref.~\onlinecite{scherer2017}, the parameters for the mean-field calculation were $ \tilde{V} = 0.74 $\,eV and $ \tilde{V}_{0} = 0 $. As can be seen from the bandstructures, SOC leads to splittings and shifts at both center and boundary of the Brillouin zone. We show bandstructures for both $ \lambda > 0 $ and $ \lambda < 0 $ where the effects of SOC are slightly different as to which type of splittings occur and in which direction states at the Brillouin zone center are shifted to. 
\begin{figure*}[h!]
\centering
\begin{minipage}{1\textwidth}
\centering
\includegraphics[width=1\textwidth]{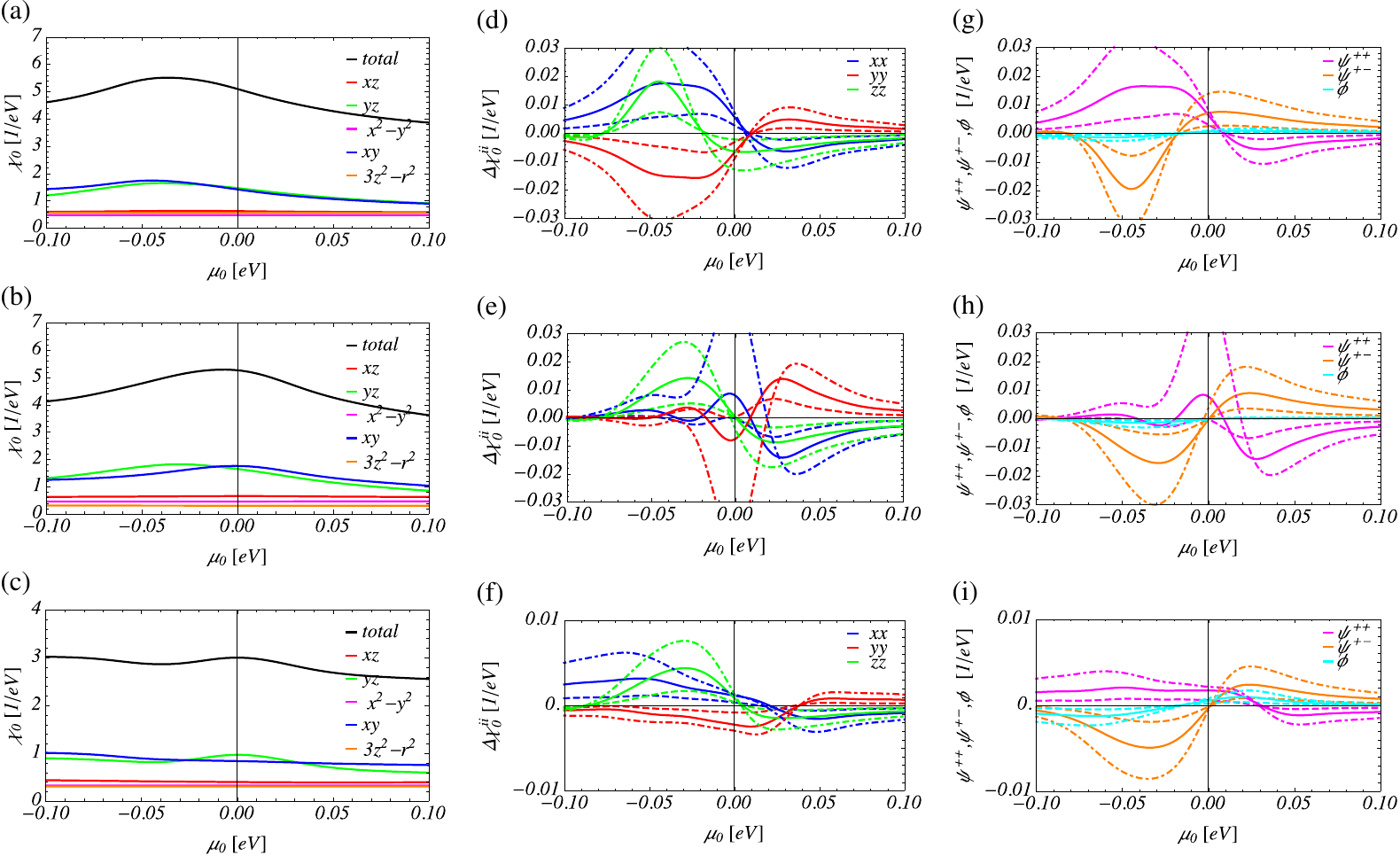}
\end{minipage}
\caption{(a),(b),(c) Chemical potential dependence of the total and orbitally ($\mu = \nu $ only) resolved isotropic contribution to the static non-interacting susceptibility with SOC $ \lambda = 0.025 $\, eV at  $ k_{\mathrm{B}} T = 0.01 $\,eV for the (a) LaFeAsO and (b) BaFe$_{2}$As$_{2}$ and (c) FeSe model with fixed wavevector ${\bf Q}_{1}$. (d),(e),(f) Corresponding anisotropic contributions and (g),(h),(i) summed particle-hole amplitudes contributing to the anisotropic magnetic response for $\lambda = 0.015$\,eV (dashed), $ \lambda = 0.025$\,eV (solid) and $ \lambda = 0.035$\,eV (dot-dashed).}
\label{fig:chi0_chemical_potential_neg_SOC}
\end{figure*}
We additionally explore the effect of a negative SOC, $\lambda < 0$, on the ansiotropy $ \Delta\chi_{0}^{ii} $, where we restrict ourselves to the same $ |\lambda | $-values as in the main text, see Fig.~\ref{fig:chi0_chemical_potential_neg_SOC}. In the LaFeAsO model, we observe a suppression of the
hole-doped $\mu_{0}$-region with dominant $\chi_{0}^{zz}$ compared to the $\lambda > 0 $ case, which can be traced back to an increase of the $ \psi^{++} $ amplitude in the corresponding doping regime. In the BaFe$_{2}$As$_{2}$ model, additional zero-crossing appear in $ \Delta \chi_{0}^{xx} $ and $ \Delta \chi_{0}^{yy} $ on the hole-doped side, while for $ \lambda = - 0.035 \,$eV the zero-crossings are removed. Both LaFeAsO and BaFe$_{2}$As$_{2}$ models show the same qualitative behavior on the electron-doped side, as they do for $ \lambda > 0 $. In the case of FeSe, the magnetic anisotropy shows the same qualitative behavior as for positive $ \lambda $ for both hole- and electron-doping. We conclude that $\mathrm{sign}(\lambda)$ can have a qualitative influence on the $ \mu_{0} $-dependence of the anisotropy, but the changes in the hierarchy of $ \chi_{0}^{ii} $ strongly depend on quantitative differences in the doping dependence of particle-hole amplitudes $\psi^{++}$, $\psi^{+-}$ and $\phi$.

\section{RPA correlation functions}
\label{sec:RPA}

Following Ref.~\onlinecite{scherer2016}, we here describe the RPA formalism we employ to analyze the collective excitations of FeSCs. As appropriate for the presence of a general SOC term, we assume the absence of spin-rotation symmetry. While in a paramagnetic state without SOC the conservation of electronic spin facilitates a decoupling of the RPA equations for transverse and longitudinal fluctuations, this is in general no longer the case in the presence of SOC. Since SOC also generates a coupling between charge- and spin-fluctuations already at the Gaussian level (in the language of effective actions for collective excitations), the RPA equations need to be extended to account for the mixing of charge- and spin-excitations for general transfer momenta. For high-symmetry momenta, like the stripe wave-vectors ${\bf Q}_{1} = (\pi,0)$ and ${\bf Q}_{2} = (0,\pi)$, the coupling between charge and spin sector vanishes. The formalism we present below is, however, general and not restricted to specific
momenta. We compute the imaginary-time spin-spin correlation function (where  $i,j$ refer to the spatial directions $x,y,z$)
\be
\chi^{ij}(\mathrm{i}\omega_n,{\bf q}) = \frac{g^2}{2}\int_{0}^{\beta} \! d\tau \,
\mathrm{e}^{\mathrm{i}\omega_n \tau}
\langle \mathcal{T}_{\tau} S^{i}_{{\bf q}}(\tau) S^{j}_{-{\bf q}}(0)\rangle,
\ee
with the Fourier transformed electron spin operator for the 2-Fe unit cell given as
\be
S^{i}_{{\bf q}}(\tau) = \frac{1}{\sqrt{\mathcal{N}}}\sum_{{\bf k},l,\mu,\sigma,\sigma^{\prime}} c_{{\bf k} - {\bf q}l\mu\sigma}^{\dagger}(\tau) \frac{\sigma_{\sigma\sigma^{\prime}}^{i}}{2}c_{{\bf k}l\mu\sigma^{\prime}}(\tau).
\ee
We note that we typically specify the transfer momentum $ {\bf q} $ with respect to the coordinate system of the 1-Fe Brillouin zone. It is then understood that `$ {\bf k} - {\bf q} $' refers to
subtraction of the two vectors in a common coordinate system. Here $\mathcal{T}_{\tau}$ denotes the time-ordering operator with respect to the imaginary-time variable $\tau \in [0,\beta)$, with $\beta$ the inverse temperature and $\sigma_{\sigma\sigma^{\prime}}^{i}$ the $i$-th Pauli matrix. From the imaginary part of $\chi^{ij}(\mathrm{i}\omega_n,{\bf q})$, we can extract the spectrum of spin-excitations that are probed by neutron scattering. The density susceptibility is defined as
\be
\chi^{00}(\mathrm{i}\omega_n,{\bf q}) = \frac{1}{2}\int_{0}^{\beta} \! d\tau \,
\mathrm{e}^{\mathrm{i}\omega_n \tau}
\langle \mathcal{T}_{\tau} N_{{\bf q}}(\tau) N_{-{\bf q}}(0)\rangle ,
\ee
with the density operator
\be
N_{{\bf q}}(\tau) = \frac{1}{\sqrt{\mathcal{N}}}\sum_{{\bf k},l,\mu,\sigma} c_{{\bf k} - {\bf q}l\mu\sigma}^{\dagger}(\tau) c_{{\bf k}l\mu\sigma}(\tau).
\ee
\begin{figure*}[t!]
\centering
\begin{minipage}{1\textwidth}
\centering
\includegraphics[width=1\textwidth]{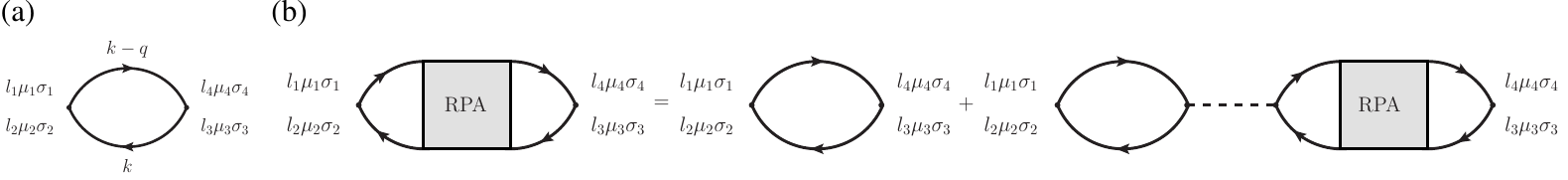}
\end{minipage}
\caption{(a) Bubble diagram for the non-interacting generalized correlation function~\Eqref{eq:genchi0}. The labels
at the vertices denote the incoming and outgoing quantum numbers $l,\mu,\sigma$ denoting sublattice, orbital and spin.
The fermionic propagator, represented by full lines with arrows, includes the effects of SOC to infinite order in the SOC strength $\lambda$.
The fermionic propagator also carries a frequency-momentum quantum number $ k = (\mathrm{i}\nu_{p},{\bf k}) $ and $ q = (\mathrm{i}\omega_{n},{\bf q})$ denotes a bosonic transfer frequency/momentum. (b) Diagrammatic
representation of the RPA equation~\Eqref{eq:RPAequation} to compute~\Eqref{eq:genchi0} within the RPA approximation. The internal quantum numbers that are summed over are not specified. The dashed horizontal line denotes the interaction vertex $ [U] $ defined in~\Eqref{eq:vertex0}-(\ref{eq:vertex3}). We note that the interaction vertex, although represented by a horizontal line, contains both direct and exchange contributions in terms of the microscopic electronic interaction.}
\label{fig:RPA_diagrams}
\end{figure*}
To derive RPA expressions for the above quantities in the absence of spin-rotation symmetry, it proves useful to introduce the generalized correlation function 
\be
\label{eq:genchi}
[\chi]^{l_1\mu_1\sigma_1;l_2\mu_2\sigma_2}_{l_3\mu_3\sigma_3; l_4\mu_4\sigma_4}(\mathrm{i}\omega_n,{\bf q})=
\frac{1}{\mathcal{N}}\int_{0}^{\beta} \! d\tau \, \mathrm{e}^{\mathrm{i}\omega_n \tau}\sum_{{\bf k},{\bf k}^{\prime}}
\langle \mathcal{T}_{\tau} 
c_{{\bf k} - {\bf q}l_1\mu_1\sigma_1}^{\dagger}(\tau) 
c_{{\bf k}l_2\mu_2\sigma_2}(\tau)
c_{{\bf k}^{\prime} + {\bf q}^{\prime}l_3\mu_3\sigma_3}^{\dagger}(0) 
c_{{\bf k}^{\prime}l_4\mu_4\sigma_4}(0)
\rangle.
\ee
To ease notation we introduce a combined index $X \equiv (l,\mu,\sigma)$ by collecting sublattice, orbital and spin indices.
In the absence of interactions, the correlation function $[\chi]^{X_1;X_2}_{X_3;X_4}(\mathrm{i}\omega_n,{\bf q}) \equiv
[\chi]^{l_1\mu_1\sigma_1;l_2\mu_2\sigma_2}_{l_3\mu_3\sigma_3; l_4\mu_4\sigma_4}(\mathrm{i}\omega_n,{\bf q}) $ reduces to
\be 
\label{eq:genchi0}
[\chi_{0}]^{X_1;X_2}_{X_3;X_4}(\mathrm{i}\omega_n,{\bf q}) & = & 
\frac{1}{\mathcal{N}}\int_{0}^{\beta} \! d\tau \, \mathrm{e}^{\mathrm{i}\omega_n \tau}\sum_{{\bf k},{\bf k}^{\prime}}
\langle \mathcal{T}_{\tau} 
c_{{\bf k} - {\bf q}X_1}^{\dagger}(\tau) 
c_{{\bf k}X_2}(\tau)
c_{{\bf k}^{\prime} + {\bf q}^{\prime}X_3}^{\dagger}(0) 
c_{{\bf k}^{\prime}X_4}(0)
\rangle_{0}  \\
& = & 
 -\frac{1}{\mathcal{N}}\int_{0}^{\beta} \! d\tau \,
\mathrm{e}^{\mathrm{i}\omega_n \tau}\sum_{X_1, \dots, X_4} \sum_{\bf k}
G_{X_2;X_3}(\tau,{\bf k})
G_{X_4;X_1}(-\tau,{\bf k}-{\bf q}) \\
& = & -\frac{1}{\beta\mathcal{N}}\sum_{p,{\bf k}}  G_{X_2;X_3}(\mathrm{i}\nu_{p},{\bf k})  G_{X_4;X_1}(\mathrm{i}\nu_{p}-\mathrm{i}\omega_{n},{\bf k}-{\bf q}) \\
& = & -\frac{1}{\mathcal{N}}\sum_{{\bf k},n_1,n_2}
[\mathcal{M}_{n_1,n_2}({\bf k},{\bf q})]^{X_1;X_2}_{X_3;X_4}
\frac{f(E_{n_1}({\bf k}-{\bf q})) - f(E_{n_2}({\bf k}))}{ {\mathrm{i}}\omega_n + E_{n_1}({\bf k}-{\bf q}) - E_{n_2}({\bf k})},
\ee
with the eigenenergies $E_{n}(\bf k)$ of the Hamiltonian $H_{0} + H_{\mathrm{SOC}}$ and $ f(\epsilon) = [\exp(\beta(\epsilon - \mu_{0})) + 1]^{-1}$ the Fermi-Dirac distribution. 
Here we defined the imaginary-time Greens function
\be
G_{X_1;X_2}(\tau,{\bf k}) & = & \langle \mathcal{T}_{\tau} 
c_{{\bf k}X_1}(\tau) 
c_{{\bf k}X_2}^{\dagger}(0)
\rangle_{0}
= 
\frac{1}{\beta} \sum_{n} \, \mathrm{e}^{-\mathrm{i}\nu_n \tau} G_{X_1;X_2}(\mathrm{i}\nu_{n},{\bf k}), 
\ee
with
\be 
 G_{X_1;X_2}(\mathrm{i}\nu_{n},{\bf k}) = \sum_{n} 
 \frac{\mathcal{U}_{{X_1,n}}({\bf k})\mathcal{U}_{{X_2,n}}^{\ast}({\bf k})}{\mathrm{i}\nu_{n} - \xi_{n}({\bf k})},
\ee
with $ \xi_{n}({\bf k}) =  E_{n}({\bf k}) - \mu_{0} $. The orbital-dressing factors entering the components of the bare correlation function read
\be 
[\mathcal{M}_{n_1,n_2}({\bf k},{\bf q})]^{X_1;X_2}_{X_3;X_4}=
\mathcal{U}_{{X_1,n_1}}^{\ast}({\bf k} - {\bf q})
\mathcal{U}_{{X_2,n_2}}({\bf k})
\mathcal{U}_{{X_3,n_2}}^{\ast}({\bf k})
\mathcal{U}_{{X_4,n_1}}({\bf k} - {\bf q}).
\ee
The unitary matrix $\mathcal{U}_{{l\mu\sigma,n}}({\bf k})$ diagonalizes the quadratic Hamiltonian $H_0 + H_{\mathrm{SOC}}$. 
The RPA equation for the generalized correlation function reads as
\be 
\label{eq:RPAequation}
[\chi]^{X_1;X_2}_{X_3; X_4}(\mathrm{i}\omega_n,{\bf q}) =
[\chi_{0}]^{X_1;X_2}_{X_3;X_4}(\mathrm{i}\omega_n,{\bf q}) +
[\chi_{0}]^{X_1;X_2}_{Y_1;Y_2}(\mathrm{i}\omega_n,{\bf q})
[U]^{Y_1;Y_2}_{Y_3;Y_4}
[\chi]^{Y_3;Y_4}_{X_3;X_4}(\mathrm{i}\omega_n,{\bf q}).
\ee
Repeated indices are summed over in \Eqref{eq:RPAequation}. The bare fluctuation vertex $[U]^{X_1;X_2}_{X_3;X_4} \equiv [U]^{l_1\mu_1\tau_{1};l_2\mu_2\tau_{2}}_{l_3\mu_3\tau_{3};l_4\mu_4\tau_{4}}$ originates from the Hubbard-Hund interaction and describes how electrons scatter off a collective excitation in the particle-hole channel. Since we employ the Hubbard-Hund interaction with interaction parameters preserving spin-rotational symmetry, it is still possible to classify the scattering of collective excitations according to their total spin. Accordingly, the vertex can be split into three different contributions as
\be 
\label{eq:vertex0}
[U]^{X_1;X_2}_{X_3;X_4} =
[U_{1}]^{X_1;X_2}_{X_3;X_4} +
[U_{2}]^{X_1;X_2}_{X_3;X_4} +
[U_{3}]^{X_1;X_2}_{X_3;X_4},
\ee
where $U_{1}$ and $U_{3}$ describe the scattering of opposite spin and equal spin fluctuations in the longitudinal channel, respectively, while $U_{2}$ describes the scattering of transverse spin fluctuations.

The vertex contribution $U_{1}$ is defined as
\be
\label{eq:vertex1}
[U_{1}]^{l\mu\sigma;l\mu\sigma}_{l\mu\bar{\sigma}; l\mu\bar{\sigma}} = - U,
\quad
[U_{1}]^{l\mu\sigma;l\mu\sigma}_{l\nu\bar{\sigma}; l\nu\bar{\sigma}} = - U^{\prime},
\quad
[U_{1}]^{l\mu\sigma;l\nu\sigma}_{l\nu\bar{\sigma}; l\mu\bar{\sigma}} = - J,
\quad
[U_{1}]^{l\mu\sigma;l\nu\sigma}_{l\mu\bar{\sigma}; l\nu\bar{\sigma}} = - J^{\prime},
\quad\text{with}\,\mu \neq \nu 
\ee
where $\bar{\sigma}$ denotes the opposite spin polarization to $\sigma$.  The $U_{1}$ contribution is zero for all other sublattice, orbital or spin index combinations. For the equal spin fluctuation vertex, we find the non-zero elements
\be
\label{eq:vertex2}
[U_{3}]^{l\mu\sigma;\mu\sigma}_{l\nu\sigma;l\nu\sigma} =  -(U^{\prime} - J),
\quad
[U_{3}]^{l\nu\sigma;\mu\sigma}_{l\mu\sigma;l\nu\sigma} = (U^{\prime}-J),
\quad\text{with}\,\mu \neq \nu.
\ee
For the transverse channel, we obtain
\be
\label{eq:vertex3}
[U_{2}]^{l\mu\bar{\sigma};\mu\sigma}_{l\mu\sigma; l\mu\bar{\sigma}} = U,
\quad
[U_{2}]^{l\nu\bar{\sigma};\mu\sigma}_{l\mu\sigma; l\nu\bar{\sigma}} = U^{\prime},
\quad
[U_{2}]^{l\nu\bar{\sigma};\nu\sigma}_{l\mu\sigma; l\mu\bar{\sigma}} = J,
\quad
[U_{2}]^{l\mu\bar{\sigma};\nu\sigma}_{l\mu\sigma; l\nu\bar{\sigma}} = J^{\prime},
\quad\text{with}\,\mu \neq \nu,
\ee
and zero else. For (residual) continuous spin-rotational symmetry, the transverse and longitudinal channels decouple and can be treated independently. We computed the non-interacting bubble with a 50 $\times$ 50 discretization for the electronic momenta in the 2D 2-Fe BZ and then solved the linear matrix equation \Eqref{eq:RPAequation} for the RPA correlation function. The exploration of 3D bandstructure effects on the magnetic anisotropy is beyond the scope of the present investigation. The RPA approximation to the spin susceptibilities $ \chi^{ij}(\mathrm{i}\omega_n,{\bf q}) $ and the density susceptibility $ \chi^{00}(\mathrm{i}\omega_n,{\bf q}) $ can be recovered by forming the appropriate linear combinations of correlation functions:
\be
\label{eq:susc_final_1}
\chi^{00}(\mathrm{i}\omega_n,{\bf q}) &  = &  
\sum_{l,l^{\prime}} \sum_{\mu,\nu} \frac{1}{2} \sum_{\sigma_{1},\dots,\sigma_{4}}
\delta_{\sigma_{1}\sigma_{2}} \delta_{\sigma_{3}\sigma_{4}}
[\chi]^{l\mu\sigma_{1};l\mu\sigma_{2}}_{l^{\prime}\nu\sigma_3; l^{\prime}\nu\sigma_4}(\mathrm{i}\omega_n,{\bf q}), \\
\label{eq:susc_final_2}
\chi^{ij}(\mathrm{i}\omega_n,{\bf q}) & = & 
\sum_{l,l^{\prime}} \sum_{\mu,\nu} \frac{g^2}{2}\sum_{\sigma_{1},\dots,\sigma_{4}}
\frac{\sigma_{\sigma_{1}\sigma_{2}}^{i}}{2} \frac{\sigma_{\sigma_{3}\sigma_{4}}^{j}}{2}
[\chi]^{l\mu\sigma_{1};l\mu\sigma_{2}}_{l^{\prime}\nu\sigma_3; l^{\prime}\nu\sigma_4}(\mathrm{i}\omega_n,{\bf q}).
\ee
The susceptibilities of the non-interacting model can be obtained in the same way by simply replacing the RPA approximation by the bare correlation function. For the sake of completeness, we specify the spin-configurations $[\sigma_{1}\sigma_{2}\sigma_{3}\sigma_{4}]$ that are summed in Eqs.~(\ref{eq:susc_final_1}),(\ref{eq:susc_final_2}) to arrive at the physical susceptibilities:
\be
\chi^{00} & : &
\,\,\,\left(+ [\uparrow\uparrow\uparrow\uparrow] + [\uparrow\uparrow\downarrow\downarrow]  +
[\downarrow\downarrow\uparrow\uparrow] + [\downarrow\downarrow\downarrow\downarrow]\right), \nn \\
\chi^{xx} & : & 
\,\,\,\left(+ [\uparrow\downarrow\uparrow\downarrow] + [\uparrow\downarrow\downarrow\uparrow]  +
[\downarrow\uparrow\uparrow\downarrow] + [\downarrow\uparrow\downarrow\uparrow]\right) \nn \\
\chi^{yy} & : & 
\,\,\,\left(- [\uparrow\downarrow\uparrow\downarrow] + [\uparrow\downarrow\downarrow\uparrow]  +
[\downarrow\uparrow\uparrow\downarrow] - [\downarrow\uparrow\downarrow\uparrow]\right) , \nn \\
\chi^{zz} & : & 
\,\,\,\left(+ [\uparrow\uparrow\uparrow\uparrow] - [\uparrow\uparrow\downarrow\downarrow]  -
[\downarrow\downarrow\uparrow\uparrow] + [\downarrow\downarrow\downarrow\downarrow]\right) , \nn \\
\chi^{xy} & : & 
\mathrm{i}\left(- [\uparrow\downarrow\uparrow\downarrow] + [\uparrow\downarrow\downarrow\uparrow] - 
[\downarrow\uparrow\uparrow\downarrow] + [\downarrow\uparrow\downarrow\uparrow]\right), \nn \\
\chi^{xz} & : & 
\,\,\,\left(+ [\uparrow\downarrow\uparrow\uparrow] - [\uparrow\downarrow\downarrow\downarrow] + 
[\downarrow\uparrow\uparrow\uparrow] - [\downarrow\uparrow\downarrow\downarrow]\right), \nn \\
\chi^{yx} & : & 
\mathrm{i}\left(-[\uparrow\downarrow\uparrow\downarrow] - [\uparrow\downarrow\downarrow\uparrow] + [\downarrow\uparrow\uparrow\downarrow] + [\downarrow\uparrow\downarrow\uparrow]\right), \nn \\
\chi^{yz} & : & 
\mathrm{i}\left(-[\uparrow\downarrow\uparrow\uparrow] + [\uparrow\downarrow\downarrow\downarrow] + [\downarrow\uparrow\uparrow\uparrow] - [\downarrow\uparrow\downarrow\downarrow]\right), \nn  \\
\chi^{zx} & : & 
\,\,\,\left(+ [\uparrow\uparrow\uparrow\downarrow ] + [\uparrow\uparrow\downarrow\uparrow] - 
[\downarrow\downarrow\uparrow\downarrow] - [\downarrow\downarrow\downarrow\uparrow]\right), \nn \\
\chi^{zy} & : & 
\mathrm{i}\left(-[\uparrow\uparrow\uparrow\downarrow] + [\uparrow\uparrow\downarrow\uparrow] + [\downarrow\downarrow\uparrow\downarrow] - [\downarrow\downarrow\downarrow\uparrow]\right). \nn
\ee
The susceptibilities $\chi^{0i}$ and  $\chi^{i0}$ that describe a response of charge (spin) due to an external field coupling linearly to spin (charge) can be obtained in a completely analogous fashion, but are not considered here. 

\section{Second-order perturbation theory in spin-orbit coupling}
\label{sec:perturbation}

%
\begin{figure*}[h!]
\centering
\begin{minipage}{1\textwidth}
\centering
\includegraphics[width=1\textwidth]{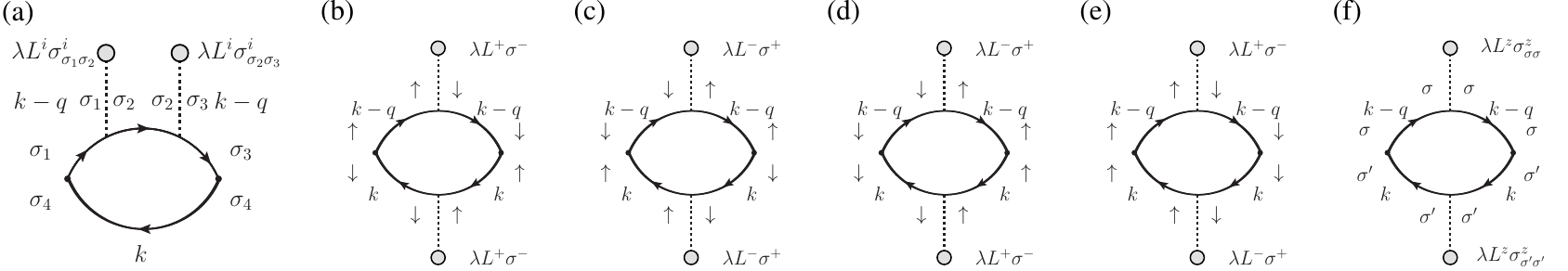}
\end{minipage}
\caption{Feynman diagrams to second order in $\lambda$ for the non-interacting susceptibility. Solid lines denote the non-interacting Greens function without SOC. While this Greens function is diagonal in the spin quantum number and both $\uparrow\uparrow$- and $\downarrow\downarrow$-components are identical in the paramagnetic state without external fields, we indicate the spin indices on the Greens function lines to show the flow of the spin quantum number through the particle-hole bubble diagrams. (a) Contribution to the isotropic part of the susceptibility, cf. \Eqref{eq:iso}. An equivalent diagram with the SOC insertions on the lower fermion line contributes as well. (b),(c) Contributions to the in-plane anisotropy, cf. Eqs.~(\ref{eq:ani1}),(\ref{eq:ani2}). (d),(e) Contributions to the out-of-plane anisotropy, cf. Eqs.~(\ref{eq:ani3}),(\ref{eq:ani5}). (f) Feynman diagram measuring the difference in the particle-hole amplitudes with parallel and antiparallel spin, cf. \Eqref{eq:ani5}.}
\label{fig:feynman_diagrams}
\end{figure*}
As another route to gain insight into the mechanism behind SOC-driven anisotropy, we 
now consider second-order perturbation theory for the non-interacting susceptibility in the
SOC strength $ \lambda $. We arrive at the following results for the components of the non-interacting Greens functions
\be 
G_{l\mu\uparrow;l^{\prime}\mu^{\prime}\uparrow}(k) & = & [G_{0}]_{l\mu;l^{\prime}\mu^{\prime}}(k)
+ \frac{\lambda}{2} [\mathcal{C}_{1}^{z}]_{l\mu;l^{\prime}\mu^{\prime}}(k)
+ \frac{\lambda^2}{4}\left(\sum_{j}[\mathcal{C}_{2}^{jj}]_{l\mu;l^{\prime}\mu^{\prime}}(k) 
+ \sum_{jk}\mathrm{i} \varepsilon^{zjk} [\mathcal{C}_{2}^{jk}]_{l\mu;l^{\prime}\mu^{\prime}}(k)\right)
+ \mathcal{O}(\lambda^{3}),
\ee
and
\be 
G_{l\mu\downarrow;l^{\prime}\mu^{\prime}\downarrow}(k) & = & [G_{0}]_{l\mu;l^{\prime}\mu^{\prime}}(k)
- \frac{\lambda}{2} [\mathcal{C}_{1}^{z}]_{l\mu;l^{\prime}\mu^{\prime}}(k)
+ \frac{\lambda^2}{4}\left(\sum_{j}[\mathcal{C}_{2}^{jj}]_{l\mu;l^{\prime}\mu^{\prime}}(k) 
- \sum_{jk}\mathrm{i} \varepsilon^{zjk} [\mathcal{C}_{2}^{jk}]_{l\mu;l^{\prime}\mu^{\prime}}(k)\right)
+ \mathcal{O}(\lambda^{3}),
\ee
as well as
\be 
G_{l\mu\downarrow;l^{\prime}\mu^{\prime}\uparrow}(k) & = &
\lambda
[\mathcal{C}_{1}^{+}]_{l\mu;l^{\prime}\mu^{\prime}}(k) + 
\mathrm{i} \frac{\lambda^2}{2}\sum_{jk} \left(\varepsilon^{xjk} + \mathrm{i} \varepsilon^{yjk} \right) [\mathcal{C}_{2}^{jk}]_{l\mu;l^{\prime}\mu^{\prime}}(k) + 
\mathcal{O}(\lambda^{3}),
\ee
and
\be 
G_{l\mu\uparrow;l^{\prime}\mu^{\prime}\downarrow}(k) & = &
\lambda
[\mathcal{C}_{1}^{-}]_{l\mu;l^{\prime}\mu^{\prime}}(k) + 
\mathrm{i}\frac{\lambda^2}{2}\sum_{jk} \left( \varepsilon^{xjk} - \mathrm{i} \varepsilon^{yjk} \right) [\mathcal{C}_{2}^{jk}]_{l\mu;l^{\prime}\mu^{\prime}}(k) + 
\mathcal{O}(\lambda^{3}),
\ee
with
\be 
[\mathcal{C}_{1}^{i}]_{l\mu;l^{\prime}\mu^{\prime}}(k) = \sum_{s}\sum_{\nu_{1},\nu_{2}}
[G_{0}]_{l\mu;s\nu_{1}}(k) [L^{i}_{s}]_{\nu_{1}\nu_{2}}  [G_{0}]_{s\nu_{2};l^{\prime}\mu^{\prime}}(k),
\ee
\be 
[\mathcal{C}_{2}^{ij}]_{l\mu;l^{\prime}\mu^{\prime}}(k) = \sum_{s,s^{\prime}}\sum_{\nu_{1} ... \nu_{4}}
[G_{0}]_{l\mu;s\nu_{1}}(k) 
[L^{i}_{s}]_{\nu_{1}\nu_{2}} 
[G_{0}]_{s\nu_{2};s^{\prime}\nu_{3}}(k)
[L^{j}_{s^{\prime}}]_{\nu_{3}\nu_{4}} 
[G_{0}]_{s^{\prime}\nu_{3};l^{\prime}\mu^{\prime}}(k),
\ee
where we defined $ [\mathcal{C}_{1}^{\pm}]_{l\mu;l^{\prime}\mu^{\prime}}(k) = \frac{1}{2}\left([\mathcal{C}_{1}^{x}]_{l\mu;l^{\prime}\mu^{\prime}}(k) \pm \mathrm{i}[\mathcal{C}_{1}^{y}]_{l\mu;l^{\prime}\mu^{\prime}}(k)\right) $ and $ G_{0} $ denotes the non-interacting Greens function without spin-orbit coupling. Defining the shorthand notation
\be 
G_{\sigma_{1}\sigma_{2}}G_{\sigma_{3}\sigma_{4}} \equiv -\frac{g^2}{4 \beta\mathcal{N}} 
\sum_{k} \sum_{l,l^{\prime}} \sum_{\mu,\mu^{\prime}}
G_{l\mu\sigma_1;l^{\prime}\nu\sigma_2}(k) 
G_{l^{\prime}\nu\sigma_3;l\mu\sigma_4}(k-q), \nn
\ee
we then arrive at
\be
\label{eq:ani1}
G_{\uparrow\downarrow}G_{\uparrow\downarrow} & = & 
 -\frac{g^2}{4 \beta \mathcal{N}} \sum_{k} \sum_{l,l^{\prime}} \sum_{\mu,\mu^{\prime}} \lambda^{2}
 [\mathcal{C}_{1}^{-}]_{l\mu;l^{\prime}\mu^{\prime}}(k) 
 [\mathcal{C}_{1}^{-}]_{l^{\prime}\mu^{\prime};l\mu}(k-q) + 
\mathcal{O}(\lambda^{3}), \\
\label{eq:ani2}
 G_{\downarrow\uparrow}G_{\downarrow\uparrow} & = & 
 -\frac{g^2}{4 \beta \mathcal{N}} \sum_{k} \sum_{l,l^{\prime}} \sum_{\mu,\mu^{\prime}} \lambda^{2}
 [\mathcal{C}_{1}^{+}]_{l\mu;l^{\prime}\mu^{\prime}}(k) 
 [\mathcal{C}_{1}^{+}]_{l^{\prime}\mu^{\prime};l\mu}(k-q) + 
\mathcal{O}(\lambda^{3}), \\
\label{eq:ani3}
G_{\uparrow\downarrow}G_{\downarrow\uparrow} & = & 
 -\frac{g^2}{4 \beta \mathcal{N}} \sum_{k} \sum_{l,l^{\prime}} \sum_{\mu,\mu^{\prime}} \lambda^{2}
 [\mathcal{C}_{1}^{-}]_{l\mu;l^{\prime}\mu^{\prime}}(k) 
 [\mathcal{C}_{1}^{+}]_{l^{\prime}\mu^{\prime};l\mu}(k-q) + 
\mathcal{O}(\lambda^{3}), \\
\label{eq:ani4}
G_{\downarrow\uparrow}G_{\uparrow\downarrow} & = & 
 -\frac{g^2}{4 \beta \mathcal{N}} \sum_{k} \sum_{l,l^{\prime}} \sum_{\mu,\mu^{\prime}} \lambda^{2}
 [\mathcal{C}_{1}^{+}]_{l\mu;l^{\prime}\mu^{\prime}}(k) 
 [\mathcal{C}_{1}^{-}]_{l^{\prime}\mu^{\prime};l\mu}(k-q) + 
\mathcal{O}(\lambda^{3}).
\ee
Thus, anisotropy in the susceptibility emerges at second order in $ \lambda $. We further obtain
\be 
\label{eq:ani5}
\frac{1}{4}\sum_{\sigma}\left[
G_{\sigma\sigma}G_{\bar{\sigma}\bar{\sigma}} -
G_{\sigma\sigma}G_{\sigma\sigma} \right] = 
-\frac{g^2}{4 \beta \mathcal{N}} \sum_{k}  \sum_{l,l^{\prime}} \sum_{\mu,\mu^{\prime}} \frac{\lambda^{2}}{4} 
[\mathcal{C}_{1}^{z}]_{l\mu;l^{\prime}\mu^{\prime}}(k) [\mathcal{C}_{1}^{z}]_{l^{\prime}\mu^{\prime};l\mu}(k-q)
+ \mathcal{O}(\lambda^{3}).
\ee
Plugging in the decomposition of the Greens function over Eigenstates of $H_{0}$, we find
(where $a$ and $b$ are the appropriate labels corresponding to the spin-combinations
 $ \sigma_{1}\bar{\sigma}_{1} $ and $ \sigma_{2}\bar{\sigma}_{2} $)
\be 
G_{\sigma_{1}\bar{\sigma}_{1}}G_{\sigma_{2}\bar{\sigma}_{2}} & = &
 -\frac{g^2}{4 \beta \mathcal{N}} \sum_{k} \sum_{l,l^{\prime}} \sum_{\mu, \mu^{\prime}} \lambda^{2}
[\mathcal{C}_{1}^{a}]_{l\mu;l^{\prime}\mu^{\prime}}(k) 
[\mathcal{C}_{1}^{b}]_{l^{\prime}\mu^{\prime};l\mu}(k-q) + \mathcal{O}(\lambda^3) \\
& = &
\frac{g^2}{4\mathcal{N}}\lambda^{2}\sum_{{\bf k}} \sum_{n_{1} ... n_{4}}
\mathcal{M}_{n_{1},n_{2};n_{3},n_{4}}({\bf k},{\bf q}) 
[L^{a}]_{n_{1}n_{2}}({\bf k})  [L^{b}]_{n_{3}n_{4}}({\bf k} - {\bf q})
\times \nn \\
& &
\mathcal{L}_{\mathrm{ani}}^{(2)}(\mathrm{i}\omega_{n};\xi_{n_{1}}({\bf k}),\xi_{n_{2}}({\bf k}),\xi_{n_{3}}({\bf k} - {\bf q}),\xi_{n_{4}}({\bf k} - {\bf q})) + \mathcal{O}(\lambda^3), \nn
\ee
and
\be  
\frac{1}{4}\sum_{\sigma}\left[G_{\sigma\sigma}G_{\bar{\sigma}\bar{\sigma}} - G_{\sigma\sigma}G_{\sigma\sigma}\right] & = &
 -\frac{g^2}{4 \beta \mathcal{N}} \sum_{k} \sum_{l,l^{\prime}} \sum_{\mu, \mu^{\prime}} \frac{\lambda^{2}}{4} 
[\mathcal{C}_{1}^{z}]_{l\mu;l^{\prime}\mu^{\prime}}(k) 
[\mathcal{C}_{1}^{z}]_{l^{\prime}\mu^{\prime};l\mu}(k-q) + \mathcal{O}(\lambda^3)
\\
& &  = 
\frac{g^2}{4\mathcal{N}}\frac{\lambda^{2}}{4}\sum_{{\bf k}} \sum_{n_{1} ... n_{4}}
\mathcal{M}_{n_{1},n_{2};n_{3},n_{4}}({\bf k},{\bf q}) 
[L^{z}]_{n_{1}n_{2}}({\bf k})  [L^{z}]_{n_{3}n_{4}}({\bf k} - {\bf q})
\times \nn \\
& & \quad
\mathcal{L}_{\mathrm{ani}}^{(2)}(\mathrm{i}\omega_{n};\xi_{n_{1}}({\bf k}),\xi_{n_{2}}({\bf k}),\xi_{n_{3}}({\bf k} - {\bf q}),\xi_{n_{4}}({\bf k} - {\bf q})) + \mathcal{O}(\lambda^3). \nn
\ee
Here we defined
\be 
\mathcal{M}_{n_{1},n_{2};n_{3},n_{4}}({\bf k},{\bf q}) \equiv \sum_{l,l^{\prime}}\sum_{\mu,\mu^{\prime}} [\mathcal{M}_{n_{1},n_{2};n_{3},n_{4}}({\bf k},{\bf q})]_{l^{\prime}\mu^{\prime};l^{\prime}\mu^{\prime}}^{l\mu;l\mu},
\ee
with the generalized product of orbital-dressing factors
\be 
[\mathcal{M}_{n_{1},n_{2};n_{3},n_{4}}({\bf k},{\bf q})]_{l_{3}\mu_{3};l_{4}\mu_{4}}^{l_{1}\mu_{1};l_{2}\mu_{2}} \equiv
\mathcal{U}_{{l_{1}\mu_{1},n_4}}^{\ast}({\bf k} - {\bf q})
\mathcal{U}_{{l_{2}\mu_{2},n_1}}({\bf k})
\mathcal{U}_{{l_{3}\mu_{3},n_2}}^{\ast}({\bf k})
\mathcal{U}_{{l_{4}\mu_{4},n_3}}({\bf k} - {\bf q}),
\ee
as well as the band-space matrix-elements of the angular momentum operator
\be 
[L^{a}]_{nn^{\prime}}({\bf k}) = 
\sum_{s}\sum_{\nu_{1},\nu_{2}}
\mathcal{U}_{s\nu_{1};n}^{\ast}({\bf k})
[L_{s}^{a}]_{\nu_{1}\nu_{2}}
\mathcal{U}_{s\nu_{2};n^{\prime}}({\bf k}).
\ee
\begin{figure*}[t!]
\centering
\begin{minipage}{0.75\textwidth}
\centering
\includegraphics[width=1\textwidth]{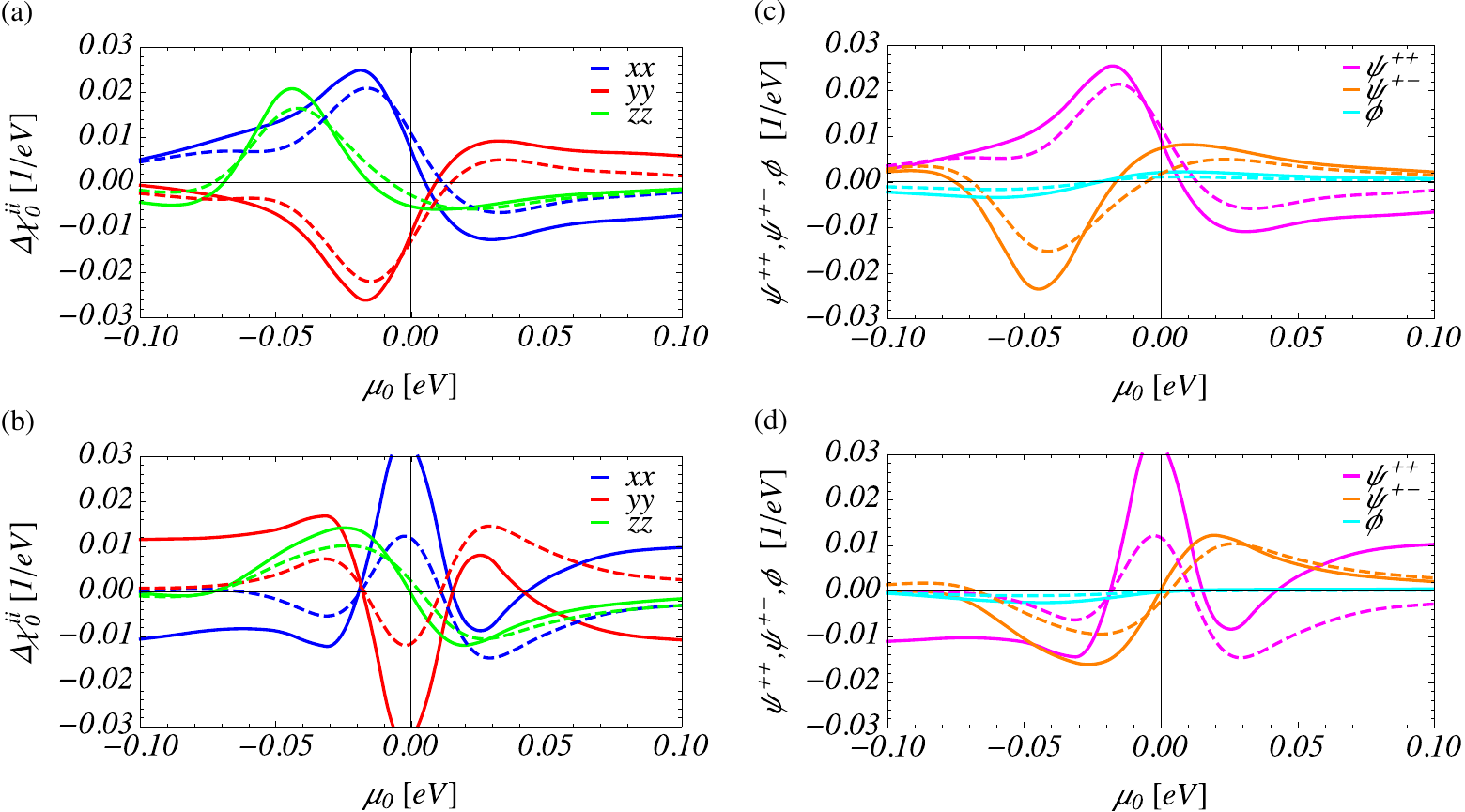}
\end{minipage}
\caption{Perturbative results (solid curves) for $\Delta\chi_{0}^{ii}$ and $\psi^{++}$, $\psi^{+-}$ and $\phi$ to order $ \lambda^{2} $ as a function of chemical potential $ \mu_{0} $ for (a),(c) LaFeAsO and (b),(d) BaFe$_{2}$As$_{2}$ models with $ \lambda = 0.025 $\,eV and $T = 0.01$\,eV compared to the exact numerical result (dashed curves).}
\label{fig:perturbation}
\end{figure*}
We further defined the Lindhard-type factor
\be 
\mathcal{L}_{\mathrm{ani}}^{(2)}(\mathrm{i}\omega_{n};\epsilon_{1},\epsilon_{2},\epsilon_{3},\epsilon_{4}) =
-\frac{1}{\beta}\sum_{p}
\prod_{j=1,2}\frac{1}{\mathrm{i}\nu_{p} - \epsilon_j}
\prod_{j^{\prime}=3,4}\frac{1}{\mathrm{i}\nu_{p} - \mathrm{i}\omega_{n} - \epsilon_{j^{\prime}}}.
\ee
For the isotropic contribution, we find
\be 
\label{eq:iso}
\frac{1}{4}\sum_{\sigma}\left[G_{\sigma\sigma}G_{\bar{\sigma}\bar{\sigma}} + G_{\sigma\sigma}G_{\sigma\sigma}\right] & = & 
-\frac{g^2}{4 \beta \mathcal{N}} \sum_{k} \sum_{l,l^{\prime}} \sum_{\mu, \mu^{\prime}}
[G_{0}]_{l\mu;l^{\prime}\mu^{\prime}}(k) 
[G_{0}]_{l^{\prime}\mu^{\prime};l\mu}(k-q)
\\
& & \hspace{-5cm} - \frac{g^2}{4 \beta \mathcal{N}} \sum_{k} \sum_{l,l^{\prime}} \sum_{\mu, \mu^{\prime}} \frac{\lambda^{2}}{4} 
\sum_{j}
\left(
[\mathcal{C}_{2}^{jj}]_{l\mu;l^{\prime}\mu^{\prime}}(k) 
[G_{0}]_{l^{\prime}\mu^{\prime};l\mu}(k-q) +
[G_{0}]_{l\mu;l^{\prime}\mu^{\prime}}(k) 
[\mathcal{C}_{2}^{jj}]_{l^{\prime}\mu^{\prime};l\mu}(k-q)  
\right)  + \mathcal{O}(\lambda^3)
\nn \\
& & \hspace{-5cm} =
\frac{g^2}{4 \mathcal{N}}\sum_{{\bf k}} \sum_{n_{1}, n_{2}}
\mathcal{M}_{n_{1},n_{2}}({\bf k},{\bf q}) 
\mathcal{L}^{(0)}(\mathrm{i}\omega_{n};\xi_{n_{1}}({\bf k}),\xi_{n_{2}}({\bf k} - {\bf q}))
\nn \\
& & \hspace{-5cm} + 
\frac{g^2}{4 \mathcal{N}}\frac{\lambda^{2}}{4}\sum_{{\bf k}} \sum_{n_{1} ... n_{4}}
\mathcal{M}_{n_{1},n_{3};n_{4},n_{4}}({\bf k},{\bf q}) 
\sum_{j}[L^{j}]_{n_{1}n_{2}}({\bf k})  [L^{j}]_{n_{2}n_{3}}({\bf k})
\times \nn \\
& & \hspace{-5cm} \quad
\mathcal{L}_{\mathrm{iso}}^{(2)}(\mathrm{i}\omega_{n};\xi_{n_{1}}({\bf k}),\xi_{n_{2}}({\bf k}),\xi_{n_{3}}({\bf k}),\xi_{n_{4}}({\bf k} - {\bf q})), \nn \\
& & \hspace{-5cm} + 
\frac{g^2}{4 \mathcal{N}}\frac{\lambda^{2}}{4}\sum_{{\bf k}} \sum_{n_{1} ... n_{4}}
\mathcal{M}_{n_{4},n_{4};n_{1},n_{3}}({\bf k},{\bf q}) 
\sum_{j}[L^{j}]_{n_{1}n_{2}}({\bf k} - {\bf q})  [L^{j}]_{n_{2}n_{3}}({\bf k} - {\bf q})
\times \nn \\
& & \hspace{-5cm} \quad
\mathcal{L}_{\mathrm{iso}}^{\prime(2)}(\mathrm{i}\omega_{n};\xi_{n_{1}}({\bf k}-{\bf q}),\xi_{n_{2}}({\bf k}-{\bf q}),\xi_{n_{3}}({\bf k}-{\bf q}),\xi_{n_{4}}({\bf k})) + \mathcal{O}(\lambda^3), \nn
\ee
with
\be 
\mathcal{M}_{n_{1},n_{2}}({\bf k},{\bf q}) \equiv \sum_{l,l^{\prime}}\sum_{\mu,\mu^{\prime}} [\mathcal{M}_{n_{1},n_{2}}({\bf k},{\bf q})]_{l^{\prime}\mu^{\prime};l^{\prime}\mu^{\prime}}^{l\mu;l\mu},
\ee
and the Lindhard-type factors
\be 
\mathcal{L}_{\mathrm{iso}}^{(2)}(\mathrm{i}\omega_{n};\epsilon_{1},\epsilon_{2},\epsilon_{3},\epsilon_{4}) & = &
-\frac{1}{\beta}\sum_{p}
\prod_{j=1,2,3}\frac{1}{\mathrm{i}\nu_{p}  - \epsilon_{j}}
\frac{1}{\mathrm{i}\nu_{p} - \mathrm{i}\omega_{n} - \epsilon_4} ,\\
\mathcal{L}_{\mathrm{iso}}^{\prime(2)}(\mathrm{i}\omega_{n};\epsilon_{1},\epsilon_{2},\epsilon_{3},\epsilon_{4}) & = &
-\frac{1}{\beta}\sum_{p}
\prod_{j=1,2,3}\frac{1}{\mathrm{i}\nu_{p} - \mathrm{i}\omega_{n} - \epsilon_{j}}
\frac{1}{\mathrm{i}\nu_{p}  - \epsilon_4}.
\ee
From these results we can read off that two SOC-insertions on one fermion line in the bubble contribute to the isotropic part, while one SOC-insertion on each fermion line generates anisotropy. The perturbative evaluation of the non-interacting susceptibility thus reveals that the leading contributions to the anisotropy in the diagonal components of the susceptibility tensor come at second order with respect to the SOC strength $ \lambda $. At $ \mathcal{O}(\lambda^3) $, the sign of the coupling $ \lambda $ can enter. As long as SOC is a perturbative scale, one can generally expect that the difference between the $ \lambda > 0 $ and $ \lambda < 0 $ cases at $ \mathcal{O}(\lambda^3) $ are most prominent, where the $ \mathcal{O}(\lambda^2) $ results indicate a change in the hierarchy of anisotropies. 

In Fig.~\ref{fig:perturbation} we compare second-order perturbation theory in SOC to the exact numerical evaluation of the non-interacting susceptibility for the chemical potential dependence of the anisotropy. As is clear from Fig.~\ref{fig:perturbation}(a),(c) the perturbation theory yields a faithful representation of the main trends observed in the full numerical result for the LaFeAsO model, while it tends to overestimate the magnitude of the anisotropic contributions. For $ \lambda $ even smaller than $ 0.025 $\,eV, the overall quantitative agreement tends to become better. For the BaFe$_{2}$As$_{2}$ model, on the other hand, qualitative agreement between full numerical evaluation and second-order perturbation theory is only found in the vicinity of the optimal nesting condition, see Fig.~\ref{fig:perturbation}(b),(d). We additionally note that the failure of perturbation theory comes mostly from the large deviations of the $\psi^{++}$-type amplitude, while both $\phi$ and $\psi^{+-}$ seem to be captured rather well by the perturbative result.

\section{Fermionic level model}
\label{app:level}

Here, we define a simple (non-interacting) model with fermionic levels 
having a well-defined orbital character in the 3$d$-manifold that are coupled
by SOC. We then evaluate the susceptibility for this simple system
and find that the anisotropy shows a behavior that is qualitatively very
similar to the variation of the spin anisotropy in the tight-binding models as a function of the chemical potential $\mu_{0}$.
To keep a certain degree of generality, we keep the five 3$d$ orbitals but do not
include a sublattice structure. It is important to note that the level structure 
in this simple Hamiltonian has nothing to do with the crystal field in the tight-binding models. The level structure rather reflects the ${\bf k}$-space nesting (`resonance' between single-particle states at different ${\bf k}$ with different orbital character).
The Hamiltonian reads
\be 
H = H_{0} + H_{\mathrm{SOC}},
\ee
with
\be 
H_{0} = \sum_{\sigma}\sum_{\mu} c_{\mu\sigma}^{\dagger} \left( \epsilon_{\mu} - \mu_{0} \right) c_{\mu\sigma}
\ee
and
\be 
H_{\mathrm{SOC}} = \frac{\lambda}{2} \sum_{\mu,\nu} \sum_{\sigma,\sigma^{\prime}}
c_{\mu \sigma}^{\dagger} [{\bf L}]_{\mu\nu}\cdot{\bf \sigma}_{\sigma\sigma^{\prime}}c_{\nu \sigma^{\prime}}.
\ee
We now compute
\be 
\chi_{0}^{ij}(\mathrm{i}\omega_{n}) & = &  
\frac{g^2}{2}\sum_{\mu,\nu}
\sum_{\sigma_{1},\dots,\sigma_{4}}
\frac{\sigma_{\sigma_{1}\sigma_{2}}^{i}}{2} \frac{\sigma_{\sigma_{3}\sigma_{4}}^{j}}{2}
[\chi_{\mathrm{0}}]^{\mu\sigma_{1};\mu\sigma_{2}}_{\nu\sigma_3;\nu\sigma_4}(\mathrm{i}\omega_n)
\\
& = & \frac{g^2}{2}\sum_{\mu,\nu} \sum_{n_{1},n_{2}}   
S_{\mu n_{1};\mu n_{2}}^{i}
S_{\nu n_{2};\nu n_{1}}^{j}
\mathcal{L}^{(0)}(\mathrm{i}\omega_{n};\epsilon_{n_1},\epsilon_{n_2}),
\ee
with the matrix elements
\be 
S_{\mu_{1} n_{1};\mu_{2} n_{2}}^{i}& \equiv & \frac{1}{2}\sum_{\sigma_{1}\sigma_{2}}
\mathcal{U}_{\mu_1\sigma_1,n_{1}}^{\ast}
\sigma^{i}_{\sigma_{1}\sigma_{2}}\mathcal{U}_{\mu_2\sigma_2,n_{2}}, \\
S_{\mu_{3} n_{3};\mu_{4} n_{4}}^{j} & \equiv & \frac{1}{2}\sum_{\sigma_{1}\sigma_{2}}
\mathcal{U}_{\mu_3\sigma_3,n_{3}}^{\ast}
\sigma^{j}_{\sigma_{3}\sigma_{4}}\mathcal{U}_{\mu_4\sigma_4,n_4},
\ee
and the Lindhard-factor
\be  
\mathcal{L}^{(0)}(\mathrm{i}\omega_{n};\epsilon_{1},\epsilon_{2}) \equiv
-\frac{1}{\beta}\sum_{p} \frac{1}{\mathrm{i}\nu_{p} - \epsilon_{1}} \frac{1}{\mathrm{i}\nu_{p} - \mathrm{i}\omega_{n} - \epsilon_{2}} =
- \frac{f(\epsilon_{1}) - f(\epsilon_{2})}{ {\mathrm{i}}\omega_n + \epsilon_{1} - \epsilon_{2}}.
\ee
The eigenstates of the Hamiltonian $ H $ obtain a non-trivial orbital structure by virtue of the
SOC that ultimately entangles the spin and orbital degree of freedom. We also note that
the only chemical potential dependence comes from the Lindhard-factor, while the orbital and spin structure of the eigenstates depends on the chosen level structure and the SOC strength $\lambda$. We also note, that while the SOC will shift the levels, it will not lead to a splitting of the originally spin-degenerate levels. So each level retains its two-fold degeneracy.

We investigate the model numerically (where we always consider the static limit $ \mathrm{i} \omega_n \to \mathrm{i}0^{+}$) and focus on two levels with $xy$ and $yz$ orbital character, respectively (the remaining levels are shifted to large negative energies). We then plot i) the isotropic contribution ii) the anisotropic contributions iii) the functions $\psi^{++}$, $\psi^{+-}$
and $\phi$ that measure different types of particle-hole excitations contributing to the anisotropy in Fig.~\ref{fig:anisotropy}, where
\be 
\psi^{++} & \equiv & +\frac{1}{2}\left[ 
G_{\uparrow\downarrow}G_{\uparrow\downarrow} +
G_{\downarrow\uparrow}G_{\downarrow\uparrow}
\right], \\
\psi^{+-} & \equiv &  +\frac{1}{2}\left[ 
G_{\uparrow\downarrow}G_{\downarrow\uparrow} +
G_{\downarrow\uparrow}G_{\uparrow\downarrow}
\right], \\
\phi & \equiv & -\frac{1}{4}\left[
G_{\uparrow\uparrow}G_{\downarrow\downarrow} +
G_{\downarrow\downarrow}G_{\uparrow\uparrow} -
G_{\downarrow\downarrow}G_{\downarrow\downarrow} -
G_{\uparrow\uparrow}G_{\uparrow\uparrow}
\right],
\ee
with the same summation conventions as defined above. For the chosen orbitals,
only the $y$-component of the angular-momentum operator has non-vanishing matrix elements. We note that while plotting all quantities in absolute units, we do not intend to relate the `physics' of this simple model to a tight-binding model.
\begin{figure*}[h!]
\centering
\begin{minipage}{1\textwidth}
\centering
\includegraphics[width=1\textwidth]{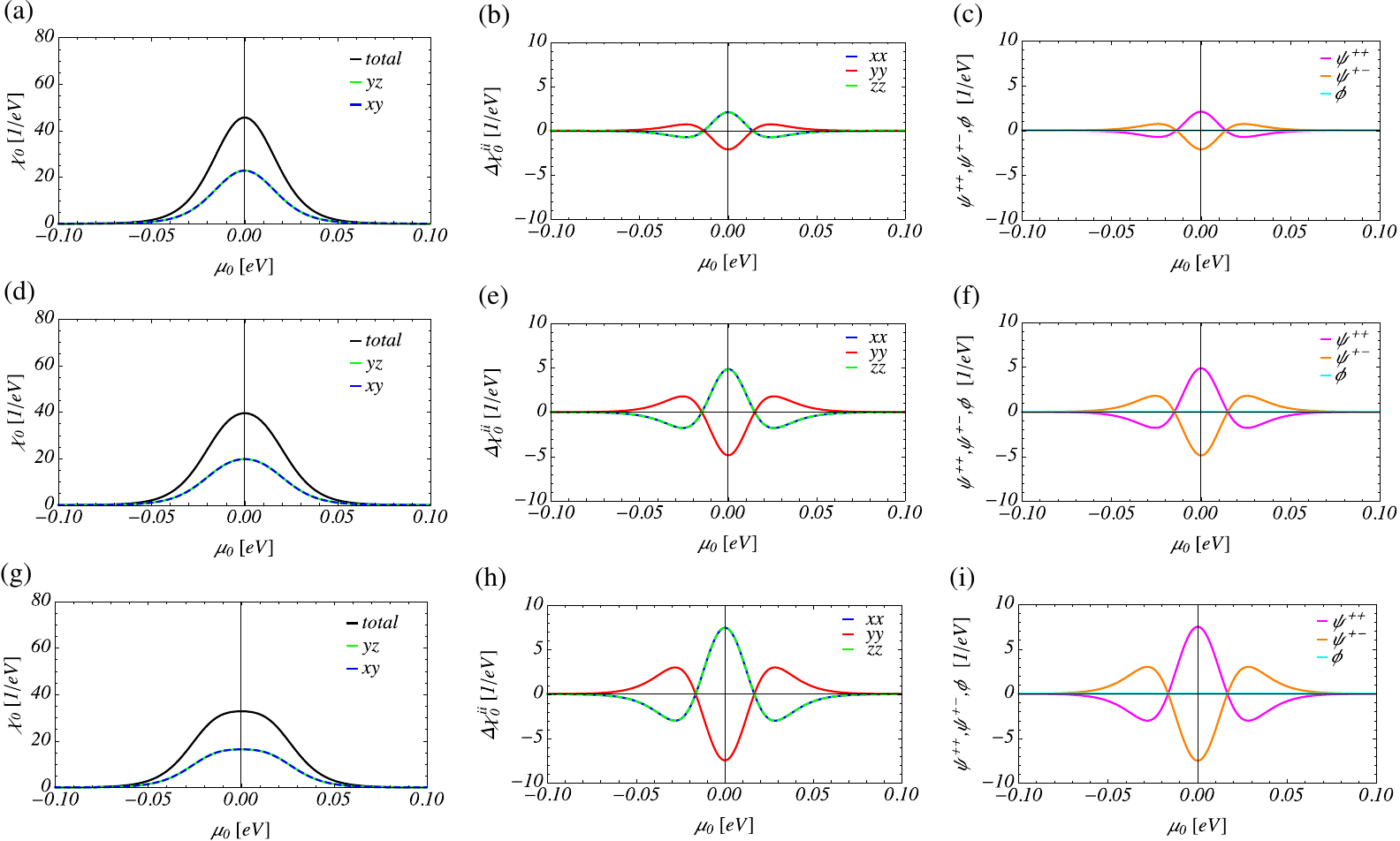}
\end{minipage}
\caption{Isotropic and anisotropic magnetic response of the simple level model with varying SOC strength, where for (a)-(c) $\lambda = 0.015$\,eV, (d)-(f) $\lambda = 0.025$\,eV, (g)-(i) $\lambda = 0.035$\,eV. (a),(d),(g) Isotropic contribution to the susceptibility. A resonance (understood as a peak in the isotropic part of the susceptibility) occurs as the chemical potential sweeps across the position of levels with $xy$ and $yz$ orbitals. The level corresponding to the remaining orbitals are shifted to large negative energies and do not contribute here. Both $xzy$ and $yz$ orbitals have the same contribution to the total susceptibility. Increasing SOC leads to a broadening of the resonance. (b),(e),(h) Chemical potential dependence of the anisotropy. The degeneracy $ \chi_{0}^{xx} = \chi_{0}^{zz} $ occurs due to the simplicity of the level model. (c),(f),(i) Particle-hole amplitudes for the level model. Also here, the relation $ \psi^{++} = - \psi^{+-} $ is due to the simplicity of the level model.}
\label{fig:anisotropy}
\end{figure*}
Rather, it serves as an analogy, that demonstrates that the qualitative behavior of the chemical potential dependence of the anisotropic response is determined to a large extent by i) the orbitals involved in the resonance and correspondingly ii) the subspace of orbitals the angular-momentum operator is acting on, as well as iii) the position of the resonance on the energy axis. It is clear from Fig.~\ref{fig:anisotropy} that including $xy$ and $yz$ type orbitals
leads to the same type of particle-hole excitations as in the tight-binding models, where $ \phi $-type excitations vanish and $ \psi^{++} $ and $ \psi^{+-} $ determine the anisotropy. In contrast to the tight-binding models, the level model produces amplitudes $ \psi^{++} $ and $ \psi^{+-} $ that obey $ \psi^{++} = - \psi^{+-} $. 
Aside from asymmetry around the position of the resonance, the qualitative $\mu_{0}$-dependence of the anisotropy response in the full model could be obtained from the level model when globally shifting the $ \psi^{+-} $ amplitude to lower energies on the energy axis. While the property  $ \psi^{++} = - \psi^{+-} $ is robust in the level model
to both level splitting and/or hybridization, bringing additional levels of different orbital character closer to the $ xy $ and $ yz $ levels can lift the degeneracy between $ \Delta \chi_{0}^{xx} $ and $ \Delta \chi_{0}^{zz} $, essentially by producing 
a finite $ \phi $ amplitude. As can be seen from Fig.~\ref{fig:anisotropy}(b),(e),(h) an increase of SOC widens the region where a particular hierarchy in the magnetic anisotropy is realized, i.e., the zero-crossings of $ \Delta\chi_{0}^{xx} $, $ \Delta\chi_{0}^{yy}$ and
$ \Delta\chi_{0}^{zz} $ move further away from the position of the peak in the isotropic part of the susceptibility. In the same way as the nesting condition in the full tight-binding model has most orbital contribution from $ xy $ and $ yz $ at ${\bf Q}_{1}$, while at $ {\bf Q}_{2} $ the dominant contributions come from $xy$ and $ xz $, replacing $ yz $ by $ xz $ changes the sign of $\psi^{++}$ and yields $ \psi^{++} = - \psi^{+-} $. In that case, $ \Delta\chi_{0}^{yy} $ and $ \Delta\chi_{0}^{zz} $ become degenerate. Aside from the degeneracy, this behavior of the anisotropy is fully in line (at least on a qualitative level) with the behavior observed in the tight-binding models, again demonstrating both the presence of a resonance and the symmetry of the involved orbitals as the deciding, universal factors. 

Albeit its simplicity, gaining a detailed understanding of the interplay of matrix elements and the Lindhard-factor seems to be involved, at least in the sense that it is an intricate interplay of inter- and intra-orbital contributions that give rise to the observed chemical potential dependence in the magnetic anisotropy. Since we do not expect these details to carry over to
the tight-binding models, we do not delve into a discussion here. While the level model clearly demonstrates universality in the chemical potential dependence of the anisotropy response, it cannot capture all the effects influencing the anisotropy in the actual tight-binding models.



\end{widetext}


\begin{thebibliography}{n}

\bibitem{dai} Pengcheng Dai, Rev. Mod. Phys. {\bf 87}, 855 (2015).

\bibitem{avci14a} S. Avci, O. Chmaissem, J. M. Allred, S. Rosenkranz, I. Eremin, A. V. Chubukov, D. E. Bugaris, D. Y. Chung, M. G. Kanatzidis, J.-P Castellan, J. A. Schlueter, H. Claus, D. D. Khalyavin, P. Manuel, A. Daoud-Aladine, and R. Osborn, Nat. Commun. {\bf 5}, 3845 (2014).

\bibitem{bohmer15a} A. E. B{\"o}hmer, F. Hardy, L. Wang, T. Wolf, P. Schweiss, and C. Meingast, Nat. Commun. {\bf 6}, 7911 (2015).

\bibitem{allred15a} J. M. Allred, S. Avci, Y. Chung, H. Claus, D. D. Khalyavin, P. Manuel, K. M. Taddei, M. G. Kanatzidis, S. Rosenkranz, R. Osborn, and O. Chmaissem, Phys. Rev. B {\bf 92}, 094515 (2015).

\bibitem{zheng16a} Y. Zheng, P. M. Tam, J. Hou, A. E. B{\"o}hmer, T. Wolf, C. Meingast, and R. Lortz, Phys. Rev. B {\bf 93}, 104516 (2016).

\bibitem{malletta} B. P. P. Mallett, Y. G. Pashkevich, A. Gusev, T. Wolf, and C. Bernhard, Euro. Phys. Lett. {\bf 111}, 57001
(2015).

\bibitem{mallettb} B. P. P. Mallett, P. Marsik, M. Yazdi-Rizi, T. Wolf, A. E. B\"{o}hmer, F. Hardy, C. Meingast, D. Munzar, and C. Bernhard, Phys. Rev. Lett. {\bf 115}, 027003 (2015).

\bibitem{allred16a} J. M. Allred, K. M. Taddei, D. E. Bugaris, M. J. Krogstad, S. H. Lapidus, D. Y. Chung, H. Claus, M. G. Kanatzidis, D. E. Brown, J. Kang, R. M. Fernandes, I. Eremin, S. Rosenkranz, O. Chmaissem, and R. Osborn, Nat. Phys. {\bf 12}, 493 (2016).

\bibitem{meier17} W. R. Meier, Q.-P. Ding, A. Kreyssig, S. L. Bud'ko, A. Sapkota, K. Kothapalli, V. Borisov, R. Valentí C. D. Batista, P. P. Orth, R. M. Fernandes, A. I. Goldman, Y. Furukawa, A. E. B\"{o}hmer, and P. C. Canfield, arXiv:1706.01067.

\bibitem{lorenzana08} J. Lorenzana, G. Seibold, C. Ortix, and M. Grilli, Phys. Rev. Lett. {\bf 101}, 186402 (2008).

\bibitem{eremin} I. Eremin and A. V. Chubukov, Phys. Rev. B {\bf 81}, 024511 (2010).

\bibitem{gastiasoro15} M. N. Gastiasoro and B. M. Andersen, Phys. Rev. B {\bf 92}, 140506(R) (2015).

\bibitem{scherer16} D. D. Scherer, I. Eremin, and B. M. Andersen, Phys. Rev. B {\bf 94}, 180405(R) (2016).

\bibitem{christensen17} M. H. Christensen, D. D. Scherer, P. Kotetes, and B. M. Andersen, Phys. Rev. B {\bf 96}, 014523 (2017).

\bibitem{wasser15} F. Wa\ss er, A. Schneidewind, Y. Sidis, S. Wurmehl, S. Aswartham, B. B\"{u}chner, and M. Braden, Phys. Rev. B {\bf 91}, 060505(R) (2015).

\bibitem{ma} Mingwei Ma, Philippe Bourges, Yvan Sidis, Yang Xu, Shiyan Li, Biaoyan Hu, Jiarui Li, Fa Wang, and Yuan Li, Phys. Rev. X {\bf 7}, 021025 (2017).

\bibitem{li} Yu Li, Weiyi Wang, Yu Song, Haoran Man, Xingye Lu, Frédéric Bourdarot, and Pengcheng Dai, Phys. Rev. B {\bf 96}, 020404(R) (2017).

\bibitem{song16} Y. Song, H. R. Man, R. Zhang, X. Y. Lu, C. L. Zhang, M.Wang, G. T. Tan, L.-P. Regnault, Y. X. Su, J. Kang, R. M. Fernandes, and P. C. Dai, Phys. Rev. B {\bf 94}, 214516 (2016).

\bibitem{johnson} P. D. Johnson, H.-B. Yang, J. D. Rameau, G.D. Gu, Z.-H. Pan, T. Valla, M. Weinert, and A.V. Fedorov, Phys. Rev. Lett. {\bf 114}, 167001 (2015).

\bibitem{watson} M. D. Watson, T. K. Kim, A. A. Haghighirad, N. R. Davies, A. McCollam, A. Narayanan, S. F. Blake, Y. L. Chen, S. Ghannadzadeh, A. J. Schofield, M. Hoesch, C. Meingast, T. Wolf, and A. I. Coldea, Phys. Rev. B {\bf 91}, 155106 (2015).

\bibitem{borisenko} S. V. Borisenko, D. V. Evtushinsky, Z.-H. Liu, I. Morozov, R. Kappenberger, S. Wurmehl, B. B\"{u}chner, A. N. Yaresko, T. K. Kim, M. Hoesch, T. Wolf, and N. D. Zhigadlo, Nat. Phys. {\bf 12}, 311 (2016).

\bibitem{hongding1} Peng Zhang, Koichiro Yaji, Takahiro Hashimoto, Yuichi Ota, Takeshi Kondo, Kozo Okazaki, Zhijun Wang, Jinsheng Wen, G. D. Gu, Hong Ding, Shik Shin, arXiv:1706.05163.

\bibitem{hongding2} Dongfei Wang, Lingyuan Kong, Peng Fan, Hui Chen, Yujie Sun, Shixuan Du, J. Schneeloch, R.D. Zhong, G.D. Gu, Liang Fu, Hong Ding, Hongjun Gao, arXiv:1706.06074.

\bibitem{qureshi2012} N. Qureshi, P. Steffens, S. Wurmehl, S. Aswartham, B. B\"{u}chner, and M. Braden, Phys. Rev. B {\bf 86}, 060410(R) (2012).

\bibitem{wang} C. Wang, R. Zhang, F.Wang, H. Luo, L. P. Regnault, P. Dai, and Y. Li, Phys. Rev. X {\bf 3}, 041036 (2013).

\bibitem{luo}H. Luo, M. Wang, C. Zhang, X. Lu, L.-P. Regnault, R. Zhang, S. Li, J. Hu, and P. Dai, Phys. Rev. Lett. {\bf 111}, 107006 (2013).

\bibitem{zhang} C. Zhang, M. Liu, Y. Su, L.-P. Regnault, M. Wang, G. Tan, Th. Br\"{u}ckel, T. Egami, and P. Dai, Phys. Rev. B {\bf 87}, 081101 (2013).

\bibitem{qureshi2014} N. Qureshi, C. H. Lee, K. Kihou, K. Schmalzl, P. Steffens, and M. Braden, Phys. Rev. B {\bf 90}, 100502 (2014).

\bibitem{matano} K. Matano, Z. Li, G. L. Sun, D. L. Sun, C. T. Lin, M. Ichioka, and Guo-qing Zheng, Europhys. Lett. {\bf 87}, 27012 (2009). 

\bibitem{song13} Yu Song, Louis-Pierre Regnault, Chenglin Zhang, Guotai Tan, Scott V. Carr, Songxue Chi, A. D. Christianson, Tao Xiang, and Pengcheng Dai, Phys. Rev. B {\bf 88}, 134512 (2013).

\bibitem{liu} Mengshu Liu, C. Lester, Jiri Kulda, Xinye Lu, Huiqian Luo, Meng Wang, S. M. Hayden, and Pengcheng Dai, PRB {\bf 85}, 214516 (2012).

\bibitem{cvetkovic13} V. Cvetkovic and O. Vafek, Phys. Rev. B {\bf 88}, 134510 (2013).

\bibitem{ikeda10} H. Ikeda, R. Arita, and J. Kune{\v s}, Phys. Rev. B {\bf 81}, 054502 (2010).

\bibitem{kopernik} H. Eschrig and K. Koepernik, Phys. Rev. B {\bf 80}, 104503
(2009).

\bibitem{scherer17} D. D. Scherer, A. Jacko, C. Friedrich, E. \c{S}a\c{s}io\u{g}lu, S. Bl\"{u}gel, R. Valent\'{i}, B. M. Andersen, Phys. Rev. B {\bf 95}, 094504 (2017).

\bibitem{christensen15} M. H. Christensen, J. Kang, B. M. Andersen, I. Eremin, and R. M. Fernandes, Phys. Rev. B {\bf 92}, 214509 (2015).

\bibitem{lipscombe} O. J. Lipscombe, Leland W. Harriger, P. G. Freeman, M. Enderle, Chenglin Zhang, Miaoying Wang, Takeshi Egami, Jiangping Hu, Tao Xiang, M. R. Norman, and Pengcheng Dai, Phys. Rev. B {\bf 82}, 064515 (2010).

\bibitem{steffens} P. Steffens, C. H. Lee, N. Qureshi, K. Kihou, A. Iyo, H. Eisaki, and M. Braden, Phys. Rev. Lett. {\bf 110}, 137001 (2013).

\bibitem{CZhang14} Chenglin Zhang, Yu Song, L.-P. Regnault, Yixi Su, M. Enderle, J. Kulda, Guotai Tan, Zachary C. Sims, Takeshi Egami, Qimiao Si, and Pengcheng Dai, Phys. Rev. B {\bf 90}, 140502(R) (2014)

\bibitem{wasser} F. Wa\ss er, C. H. Lee, K. Kihou, P. Steffens, K. Schmalzl, N. Qureshi, and M. Braden, Scientific Reports {\bf 7}, 10307 (2017).

\bibitem{sm}
Supplementary material.

\end{thebibliography}

\begin{thebibliography}{00}

\bibitem{ikeda2010}
H. Ikeda, R. Arita, and J. Kune$\check{\mathrm{s}}$, 
Phys. Rev. B {\bf 81}, 054502 (2010).

\bibitem{eschrig2009}
H. Eschrig and K. Koepernik,
Phys. Rev. B {\bf 80}, 104503 (2009).

\bibitem{scherer2017} D. D. Scherer, A. Jacko, C. Friedrich, E. \c{S}a\c{s}io\u{g}lu, S. Bl\"{u}gel, R. Valent\'{i}, B. M. Andersen, Phys. Rev. B {\bf 95}, 094504 (2017).

\bibitem{scherer2016}
D. D. Scherer, I. Eremin, and B. M. Andersen,
Phys. Rev. B {\bf 94}, 180405(R) (2016).

\end{thebibliography}
\end{document}